\documentclass[journal=jctcce,manuscript=article]{achemso}
\usepackage{graphicx}
\usepackage{amsmath}
\usepackage{wrapfig}
\usepackage{physics}
\usepackage{caption}
\usepackage{hyperref}
\usepackage{subfiles}
\usepackage{longtable}

\usepackage{multicol}

\graphicspath{ {figures/} }

\title{Probing the force field sensitivity of entropy and enthalpy differences in organic polymorphs using classical potentials}
\author{Nathan S. Abraham}
\affiliation{Department of Chemical and Biological Engineering, University of Colorado Boulder, Boulder, CO 80309, USA}
\author{Marcus T. Hock}
\affiliation{Department of Chemical and Biological Engineering, University of Colorado Boulder, Boulder, CO 80309, USA}
\author{Michael R. Shirts}
\affiliation{Department of Chemical and Biological Engineering, University of Colorado Boulder, Boulder, CO 80309, USA}
\email{michael.shirts@colorado.edu}

\begin{document}
\begin{singlespace}
\begin{multicols}{2}

\maketitle
\bibliographystyle{plain}

\section{Abstract}
	We evaluate the effectiveness of different classical potentials to predict the thermodynamics of a number of organic solid form polymorphs relative to experimentally reported values using the quasi-harmonic approximation. Using the polarizable potential AMOEBA we are able to predict the correct sign of the enthalpy difference for 71$\pm$12 \% of the polymorphs. Alternatively, all point charge potentials perform on par with random chance of correcting the correct sign (50\%) for enthalpy. We find that the entropy is less sensitive to the accuracy of the potential with all force fields, excluding CGenFF, reporting the correct sign of the entropy for 64$\pm$13 -- 75$\pm$11 \% of the systems. Predicting the correct sign of the enthalpy and entropy differences can help indicate the low and high temperature stability of the polymorphs, unfortunately the error relative to experiment in these predicted values can be as large as 1--2.5 kcal/mol at the transition temperature. 

\section{Introduction}
    Theoretical predictions of crystal polymorph stability are aimed at determining the crystal energy landscape of a given molecule to help aid in material design. Organics commonly pack in multiple stable and metastable forms, which can alter the solid form properties.~\cite{brittainPolymorphismPharmaceuticalSolids2016,censiPolymorphImpactBioavailability2015,chemburkarDealingImpactRitonavir2000b,chithambararajHydrothermallySynthesizedHMoO32016,fabbianiHighpressureStudiesPharmaceutical2006,giriTuningChargeTransport2011a,haasHighChargecarrierMobility2007a,haleblianPharmaceuticalApplicationsPolymorphism1969,higuchiPolymorphismDrugAvailability1963,llinasPolymorphControlPresent2008,millarCrystalEngineeringEnergetic2012a,millerIdentifyingStablePolymorph2005,milosovichDeterminationSolubilityMetastable1964,romeroSolubilityBehaviorPolymorphs1999,singhalDrugPolymorphismDosage2004a,sniderPolymorphismGenericDrug2004,stevensTemperatureMediatedPolymorphismMolecular2015a,valleOrganicSemiconductorsPolymorphism2004a,vanderheijdenCrystallizationCharacterizationRDX2004b,yuanUltrahighMobilityTransparent2014a,yuScientificConsiderationsPharmaceutical2003} Polymorphism can help aid in material design, but the emergence of previously unknown polymorphs has the potential to be detrimental to the development of pharmaceuticals.~\cite{bauerRitonavirExtraordinaryExample2001a,chemburkarDealingImpactRitonavir2000b,millerIdentifyingStablePolymorph2005} One popular class of methods to determine the crystal energy landscape is through crystal structure predictions (CSPs). CSPs exhaustively generate potential crystal structure and determine a relative stability of the crystals with only prior knowledge of the molecular chemical structure. Historically, the most common way to rank crystal structures is based on their lattice energies and the lowest energy structures, those closest to the global minimum, are assumed to exist experimentally.

    Getting a correct lattice energy ranking is largely dependent on the potential used to describe the intermolecular interactions. Early work by Day \textit{et al.} looked at CSPs of 50 rigid molecules comparing the use of fixed charges to multipoles when determining the crystals electrostatic interactions. The success of using atomic multipoles was summarized by all experimental structures being within 1.22 kcal/mol of the global minimum rather than 1.74 kcal/mol for point charge potentials. By using atomic multipoles the known experimental structures within the CSP were closer in energy and ranking to the global minimum than the results produced using fixed charges. Despite the improvement, the study concluded found that neither electrostatic model were ideal, especially for crystals with hydrogen bonding.~\cite{dayIsotropicAtomModel2005c} Since that study in 2005 a lot of focus has been placed on developing and implementing potentials with more realistic representations of the crystal, potentials that could more accurately describe interacts similar to hydrogen bonding. To accurately model these interactions for the lattice energy quantum mechanical approaches are generally needed. So in a CSP, it is common now to perform energy minimizations and energy rankings with a classical potential and then ``polish'' the final ranking using quantum mechanical  approaches.~\cite{habgoodTestingVarietyElectronicStructureBased2011,johanssonRevisionCrystalStructure2016,karamertzanisCanFormationPharmaceutical2009,nymanAccuracyReproducibilityCrystal2019,reillyReportSixthBlind2016,shtukenbergPowderDiffractionCrystal2017} The lattice energy is important for determine the relative stability of crystals, but these approaches fail to address the importance of entropy.

    Free energy stabilities give a better thermodynamic description of the crystal lattice, which is why methods to approximate entropic stability have become more prevalent in CSPs. A challenge to the utility of lattice-energy based CSPs are enantiotropic polymorphs, which change in stability ordering with temperature. While there are many cases of monotropic crystal polymorphs,~\cite{perlovichPolymorphismMonotropicForms2020} the presence of enatiotropic systems is far too prevalent to ignore when conducting a CSP.~\cite{alcobeTemperatureDependentStructuralProperties1994,badeaFusionSolidtosolidTransitions2007a,boldyrevaEffectHighPressure2002a,boldyrevaHighpressureDiffractionStudies2008,cansellPhaseTransitionsChemical1993a,cesaroThermodynamicPropertiesCaffeine1980,cherukuvadaPyrazinamidePolymorphsRelative2010b,grzesiakComparisonFourAnhydrous2003a,hasegawaReevaluationSolubilityTolbutamide2009,maherSolubilityMetastablePolymorph2012a,sacchettiThermodynamicAnalysisDSC2000a,seryotkinHighpressurePolymorphChlorpropamide2013a,stolarSolidStateChemistryPolymorphism2016b,torrisiSolidPhasesCyclopentane2008a,vemavarapuCrystalDopingAided2002,yoshinoContributionHydrogenBonds1999b,yuMeasuringFreeEnergyDifference2005a} Nyman and Day evaluated the polymorph free energy difference of over 500 rigid molecules at 0 K and close to their melting temperature using the harmonic phonons to determine the entropic contribution to the free energy difference. Of the systems, 21 \% were shown to re-rank with temperature, highlighting the importance of entropy in determining the polymorph stability.~\cite{nymanModellingTemperaturedependentProperties2016,nymanStaticLatticeVibrational2015a} This is certainly an underestimate for reranking of flexible molecules where the additional degrees of freedom will result in greater entropic contributions. Predictions can also improve the ranking by determining the free energy differences of the crystals, which can either be done with use of the harmonic phonons to approximate the entropic contribution or with molecular dynamics to determine the exact free energy ranking.~\cite{nymanAccuracyReproducibilityCrystal2019,reillyReportSixthBlind2016,schneiderExploringPolymorphismBenzene2016,shtukenbergPowderDiffractionCrystal2017}

    Historically, CSPs have primarily focused on how force fields improve the relative lattice ranking, but will now need to understand the affect on free energy. It is clear that potentials that more accurately describe the molecular/atomic interactions improve both the lattice stability and the overall prediction accuracy for a CSP.  Current research has left unclear the sensitivity of entropy as a function of the force field. There are a number of approximate and exact approaches to determine the free energy differences between polymorphs. These approaches have shown to accurately determine  solubility,~\cite{palmerFirstPrinciplesCalculationIntrinsic2012} entropy,~\cite{abrahamAddingAnisotropyStandard2019,brandenburgThermalExpansionCarbamazepine2017a,dybeckCapturingEntropicContributions2017,dybeckEffectsMoreAccurate2016a,erbaThermalPropertiesMolecular2016a,heitHowImportantThermal2016,valleOrganicSemiconductorsPolymorphism2004a} melting temperature,~\cite{zhangFullySilicoMelting2013} and thermal expansion~\cite{abrahamAddingAnisotropyStandard2019,brandenburgThermalExpansionCarbamazepine2017a,erbaThermalPropertiesMolecular2016a} relative to experiment. The success of determining these properties are independent of the potential used, leading us to hypothesize that entropic may be less sensitive to the potential used than the lattice energy. Here, we use 5 classical potentials with polarizable or fixed charge electrostatics and determine how entropy and enthalpy of enantiotropic polymorphs are affected by the different potentials.

\section{Methods}
\end{multicols}
\begin{figure}
  \includegraphics[width=16cm]{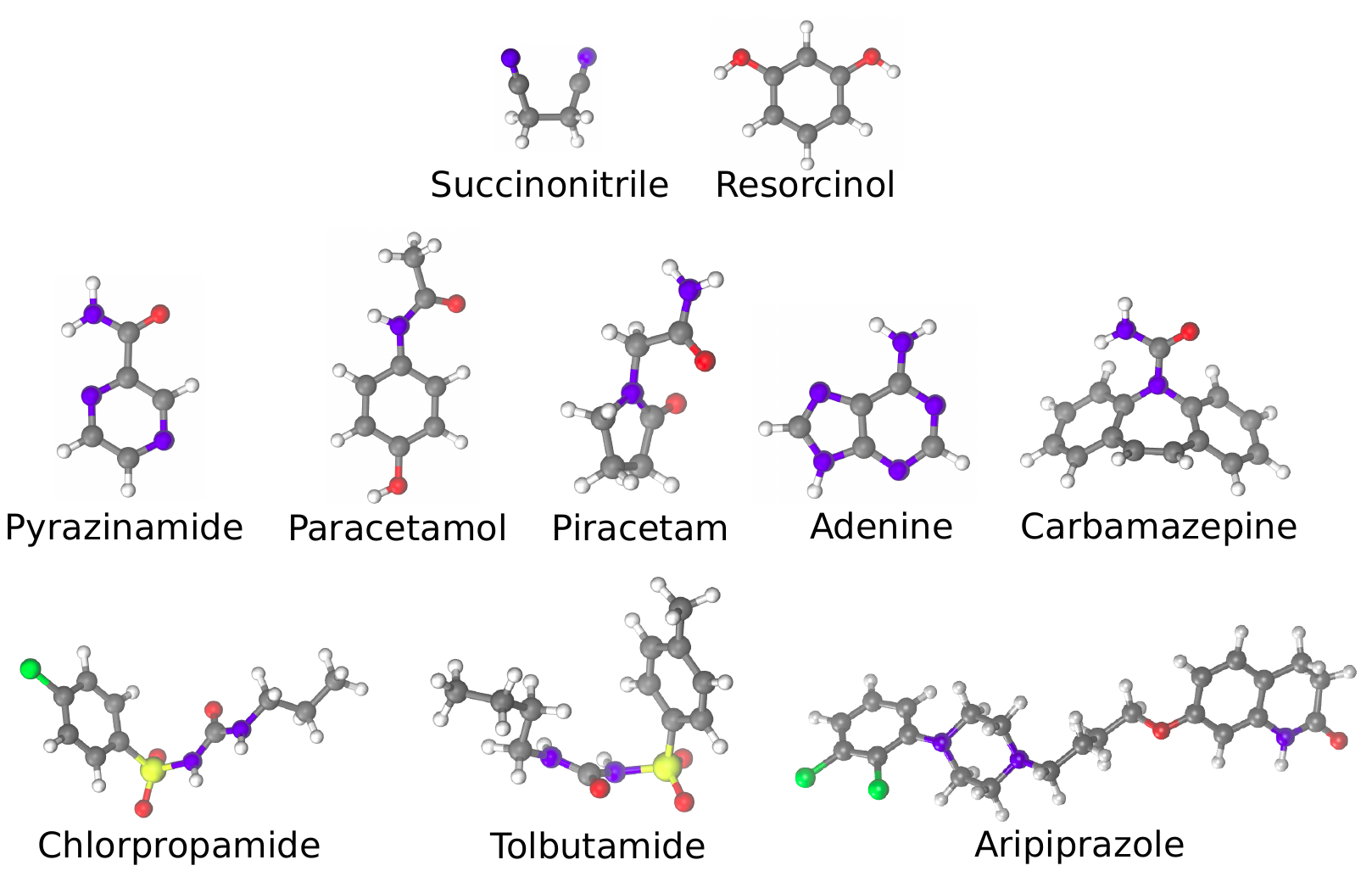}
  \caption{Molecules considered in this study.
  \label{figure:molecules}}
\end{figure}
\begin{multicols}{2}

    We examine the stability sensitivity of the 10 enantiotropic polymorphs systems in figure ~\ref{figure:molecules}, a number of which have been studied in previous publications of ours.~\cite{abrahamAddingAnisotropyStandard2019,dybeckCapturingEntropicContributions2017} The corresponding CCDC refcodes and supercell sizes used can be found in the Supporting Information (Table S\ref{table:systems}).

\subsection{Off-the-Shelf Potentials}
    We will compare the sensitivity in crystalline thermodynamics of four off-the-shelf point-charge potentials and one polarizable potential. Each molecule in figure ~\ref{figure:molecules} was parameterized using the following five potentials: 
  \begin{itemize}
    \item OPLS2 parameterized using Maestro;~\cite{wangAccurateReliablePrediction2015} 
    \item GAFF parameterized using AmberTools with AM1BCC charges generated with OpenEye;~\cite{jakalianFastEfficientGeneration2000,wangDevelopmentTestingGeneral2004} 
    \item SMIRNOFF parameterized using OpenFF with AM1BCC charges generated with OpenEye;~\cite{jakalianFastEfficientGeneration2000,mobleyEscapingAtomTypes2018}  
    \item CGenFF parameterized using cgen-ff;~\cite{vanommeslaegheCHARMMGeneralForce2010} and
    \item AMOEBA (eq~\ref{eq:U_4}).~\cite{ponderCurrentStatusAMOEBA2010}
  \end{itemize}

These four force fields all differ in the properties they seek to optimize for their parameter sets, but all have very similar functional forms with slight differences that will not be discussed here. The general functional form of the fixed charge potentials is reported in equaiton ~\ref{eq:fixed_potential}.

 \end{multicols}
    \begin{eqnarray}
    U_{total} &=& \sum \limits^{bonds}  k^{b}_{ij} (r_{ij} - r^{0}_{ij})^{2} \nonumber \\
              &+& \sum \limits^{angles} k^{\theta}_{ijk} (\theta_{ijk} - \theta^{0}_{ijk})^{2} \nonumber \\
              &+& U_{dihedrals} \nonumber \\
              &+& U_{misc} \nonumber \\
              &+& \sum \limits_{i > j}^{nonbondeds} f_{ij} 4 \epsilon_{ij} \left( \left(\frac{\sigma_{ij}}{r_{ij}}\right)^{12} - 2\left(\frac{\sigma_{ij}}{r_{ij}}\right)^{6} + \frac{q_{i} q_{j} e^{2}}{4\pi \epsilon_{0} r_{ij}} \right)     
  \end{eqnarray} \label{eq:fixed_potential}
\begin{multicols}{2}

Where $k_{b}$ is the bond force constant, $r$ and $r_{0}$ are the current and equilibrium bond lengths, $k_{\theta}$ is the angle force constant, $\theta$ and $\theta_{0}$ are the current and equilibrium angles, $f_{ij}$ is the fudge factor between atoms $i$ and $j$, $\epsilon_{ij}$ and $\sigma_{ij}$ are the Lennard-Jones potential well depth and equilibrium distances for computing the van der Waals interactions, $r_{ij}$ is the distance between atom $i$ and $j$, and $q_{i}$ is the point charge on atom $i$. 

    The energy for $U_{dihedral}$ and $U_{misc}$ vary between the four potentials. For OPLS2 $U_{dihedral}$ uses Fourier dihedrals for both proper and improper dihedrals. GAFF and SMIRNOFF $U_{dihedral}$ uses the proper dihedral form ($\sum \limits_4 \frac{F_{n}}{2}\left( 1 + \cos{(n\phi)} \right)$) for both proper and improper dihedrals. OPLS2, GAFF, and SMIRNOFF all have no additional parameters, so $U_{misc}=0$. CGenFF uses proper dihedral form for proper dihedrals, harmonic dihedrals for impropers ($k_{\phi}(\phi - \phi_{0})^{2}$), and also has a mixed bond--angle term ($k^{b\theta}_{ijk} (r_{ik} - r_{ik}^{0})^{2}$). The combining rules for the LJ parameters are:
  \begin{itemize}
    \item OPLS2 geometric for both $\sigma$ and $\epsilon$;
    \item GAFF arithmetic for $\sigma$ and geometric for $\epsilon$;
    \item SMIRNOFF arithmetic for $\sigma$ and geometric for $\epsilon$; and
    \item CGenFF arithmetic for $\sigma$ and geometric for $\epsilon$.
  \end{itemize}

    The potential for AMEOBA varies further from the fixed charge potential, with a couple of additional terms. The functional form for AMOEBA is shown in equation ~\ref{eq:U_4}.
\end{multicols}
  \begin{eqnarray}
    U_{total} &=& \sum \limits^{bonds}  k^{b}_{ij} (r_{ij} - r^{0}_{ij})^{2} \nonumber \\
              &+& \sum \limits^{angles} k^{\theta}_{ijk} (\theta_{ijk} - \theta^{0}_{ijk})^{2} \nonumber \\
              &+& \sum \limits^{bond-angle} k_{b\theta}\left((r_{ij} - r_{ij}^{0}) + (r_{jk} - r_{jk}^{0})\right)(\theta_{ijk} - \theta_{ijk}^{0}) \nonumber \\
              &+& \sum \limits^{out-of-plane} k_{\chi} \chi^{2} \nonumber \\
              &+& \sum \limits^{dihedrals} \frac{1}{2}\left(F_{1}(1+\cos{(\phi)} + F_{2}(1-\cos{(2\phi)} +F_{3}(1+\cos{(3\phi)} + F_{4}(1-\cos{(4\phi)}) \right) \nonumber \\
              &+& \sum \limits^{nonbonded} U_{vdW} + U_{multipoles} + U_{polarization}
  \end{eqnarray} \label{eq:U_4}
\begin{multicols}{2}

Where the first three summations are similar to fixed charge potentials, $k_{\chi}$ is the force constant and $\chi$ is the out of plane angle for sp2 hybridized groups, and the dihedrals are the same as OPLS2. For the non-bonded interactions there are some large differences. With the electrostatics being modeled with permanent and induced multipoles. $U_{vdW}$ uses a buffered 14--7 function with the combining rules:
\begin{eqnarray}
  \epsilon_{ij} &=& \frac{4 \epsilon_{i} \epsilon_{j}}{(\epsilon_{i}^{0.5} + \epsilon_{j}^{0.5})^{2}} \\
  \sigma_{ij} &=& \frac{\sigma_{i}^{3} + \sigma_{j}^{3}}{\sigma_{i}^{2} + \sigma_{j}^{2}}
\end{eqnarray}

\subsection{The theoretical thermodynamics were determined using the (quasi-)harmonic approximation}
    The quasi-harmonic approximation assumes that the entropic contributions of the crystal are a sum of the harmonic static lattice modes for the lattice geometry that minimizes the free energy at a given temperature and pressure. Since we are using classical potentials we will be using the classical limit for the Helmholtz free energy of a harmonic oscillator ($A_{v}$).
    \begin{eqnarray}
      G(T,P) &=& \min_{\boldsymbol{C}} f(\boldsymbol{C},T,P) \label{eq:QHA_min}  \\
      f(\boldsymbol{C},T,P) &=& \min_{\boldsymbol{x}} \left(U(\boldsymbol{C},\boldsymbol{x})\right) + A_{v}(\boldsymbol{C},T) + PV(\boldsymbol{C}) \label{eq:QHA} \\
      A_{v}(\boldsymbol{C},T) &=& \sum_{k} \beta^{-1} \ln{\left(\beta \hbar \omega_{k}(\boldsymbol{C})\right)} \label{eq:classical}
    \end{eqnarray}
Where the Gibbs free energy ($G$) at a given temperature ($T$) and pressure ($P$) is determined by finding the lattice geometry ($\boldsymbol{C}$) that minimized the free energy. The free energy is a sum of the potential energy ($U$), Helmholtz free energy of a harmonic oscillator ($A_{v}$), and a $PV$ term, which is small at the pressures we're working at. In eq.~\ref{eq:classical} the Helmholtz free energy is a sum of the energy attributed to each vibrational frequency ($\omega_{k}$). We can simplify QHA further by using the Gr\"{u}neisen parameter ($\gamma_{k,i}$) in eq~\ref{eq:aniso_Gru}, which approximates the change in the $k^{th}$ vibrational mode due to changes in any of the six principle lattice strains ($\eta_i$).
    \begin{eqnarray}
      \gamma_{k,i} = - \left.\frac{1}{\omega_k} \frac{\partial \omega_k}{\partial \eta_i}\right|_{\eta_j \ne \eta_i} \label{eq:aniso_Gru}
    \end{eqnarray}

    Our recent development of a gradient approach in conjunction with a variant of anisotropic expansion allows us to determine the free energy minimum for QHA with little error. In previous work we presented a 1D-variant to anisotropic expansion, which assumed the ratio of thermal expansion between the lattice parameters remains constant with temperature.~\cite{abrahamAddingAnisotropyStandard2019,abrahamThermalGradientApproach2018a} We found that this variant to anisotropic expansion introduced error less than 0.01 kcal/mol to the computed polymorph free energy differences and therefore will exclusively use this approach in this paper.~\cite{abrahamAddingAnisotropyStandard2019}

\subsection{Simulation Details}
    Temperature replica exchange was performed to overcome crystal restructuring that standard QHA cannot. All crystals in the four fixed charge potentials were simulated with temperature REMD for a temperature range between 10 -- 400 K with replica spacing that achieved approximately a 0.3 probability of exchange per replica. All crystals were run for 20 ns with a time step of 0.0005 ps. All simulations were performed with Gromacs 2019.3.

    The quasi-harmonic approximation was performed on the lattice minimum structure found by energy minimizing the 10 K replica of temperature REMD. All harmonic approaches were run using the Tinker molecular modeling package 8.7. For the fixed charge potentials, the lattice minimum structure was found by energy minimizing 5 random configurations from the equilibrated 10 K NPT simulation from REMD. Here the lattice minimum were found using \textit{xtalmin} to a tolerance of 10$^{-5}$ and the lowest energy minimum was selected for QHA. QHA was performed using our Python wrapper package that is available on GitHub at \url{http://github.com/shirtsgroup/Lattice_dynamics}.~\cite{abrahamThermalGradientApproach2018a}

    Restructuring is not considered for crystals parameterized with AMOEBA. AMOEBA is currently implemented in Tinker and Force Field X, but neither have temperature REMD implemented. Each lattice minimum was determined by energy minimizing the experimental structure directly from the CCDC. Using \textit{xtalmin} the experimental structures were energy minimized to a tolerance of 10$^{-5}$. We then performed HA on all crystals using our Python wrapper package \url{Lattice_dynamics}.

\section{Results and Discussion}
\begin{figure*}
  \includegraphics[width=16cm]{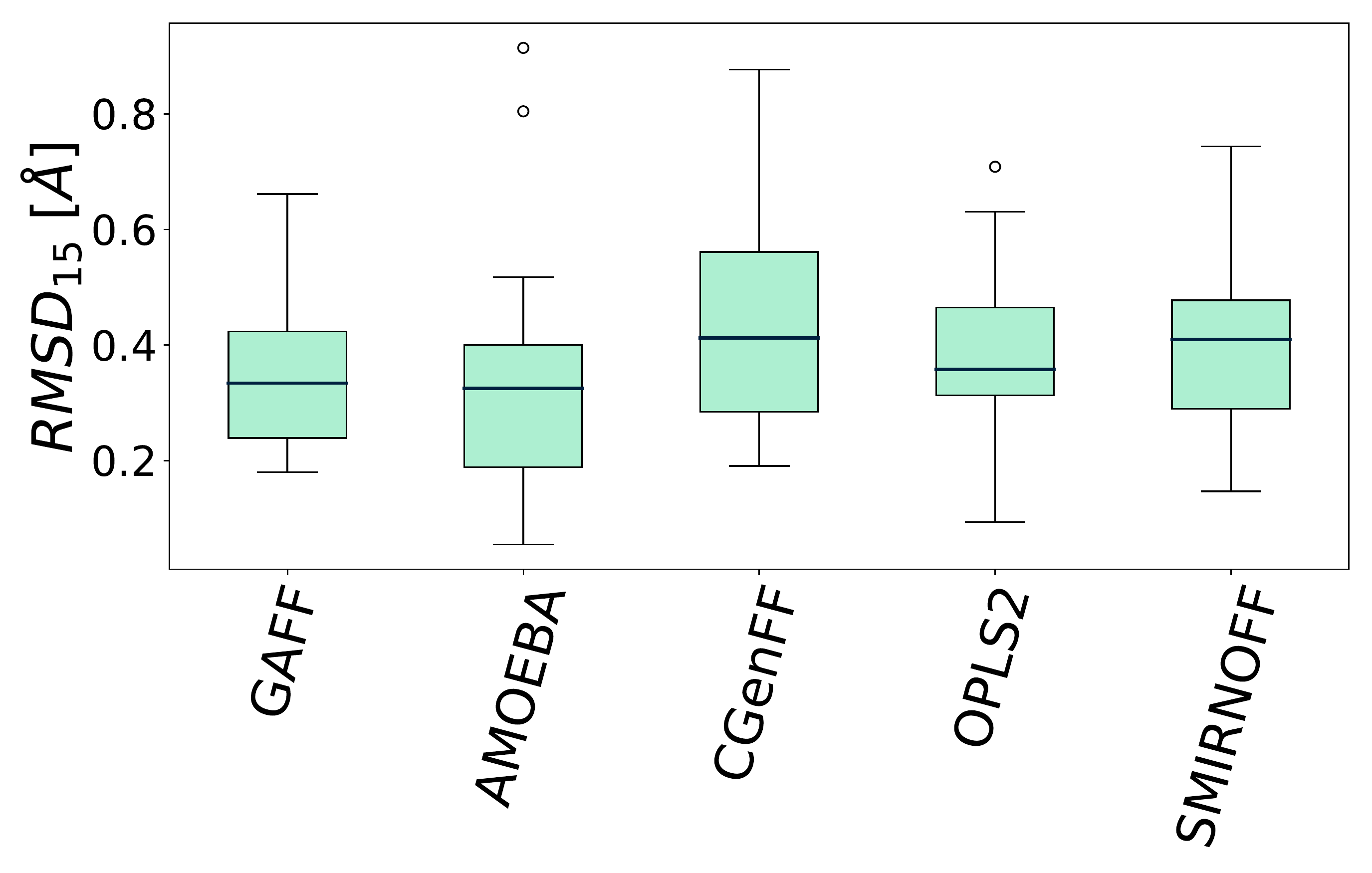} 
  \caption{The RMSD$_{15}$ of the lattice minimum structure with the experimental structure are reported in the box plots. All of the force fields perform similar to one another for the 2$^{nd}$ and 3$^{rd}$ quartile, with CGenFF performing the worst in the 4$^{th}$ quartile and AMOEBA having 3 serious outliers. Crystollographically, an RMSD$_{15}$ between crystals that is $<$ 0.3--0.5 \AA~ conventionally indicates that the two crystals belong to the same minimum.
  \label{figure:RMSD}}
\end{figure*}

    All force fields are able to find a lattice minimum structurally similar to the experimental crystal,  with AMOEBA possibly performing slightly better. In figure ~\ref{figure:RMSD} box plots for the RMSD$_{15}$ of the lattice minimum structure in each potential relative to the experimental structure are reported. The RMSD$_{15}$ is computed by taking the RMSD$_{15}$ between clusters of 15 molecules in each crystal. A cluster of 15 nearest molecules is a sufficient size to get a similarity measurement between crystals.~\cite{chisholmCOMPACKProgramIdentifying2005} CSPs generally classify crystals with RMSD$_{15}$ $<$ 0.3--0.5 \AA~as belonging to the same minimum. The majority of the lattice minimum observed in all 5 potentials are within 0.5 \AA~of the experimental structures. 
Despite two outliers, piracetam form I and tolbutamide form I, the RMSD$_{15}$ for AMOEBA are slightly smaller than the four fixed charge potentials. Polarization is known to improve lattice energies relative to fixed charge potentials and it is logical to expect that a more accurate lattice minimum would lead to a better lattice energy ranking.

\begin{figure*}
  \includegraphics[width=16cm]{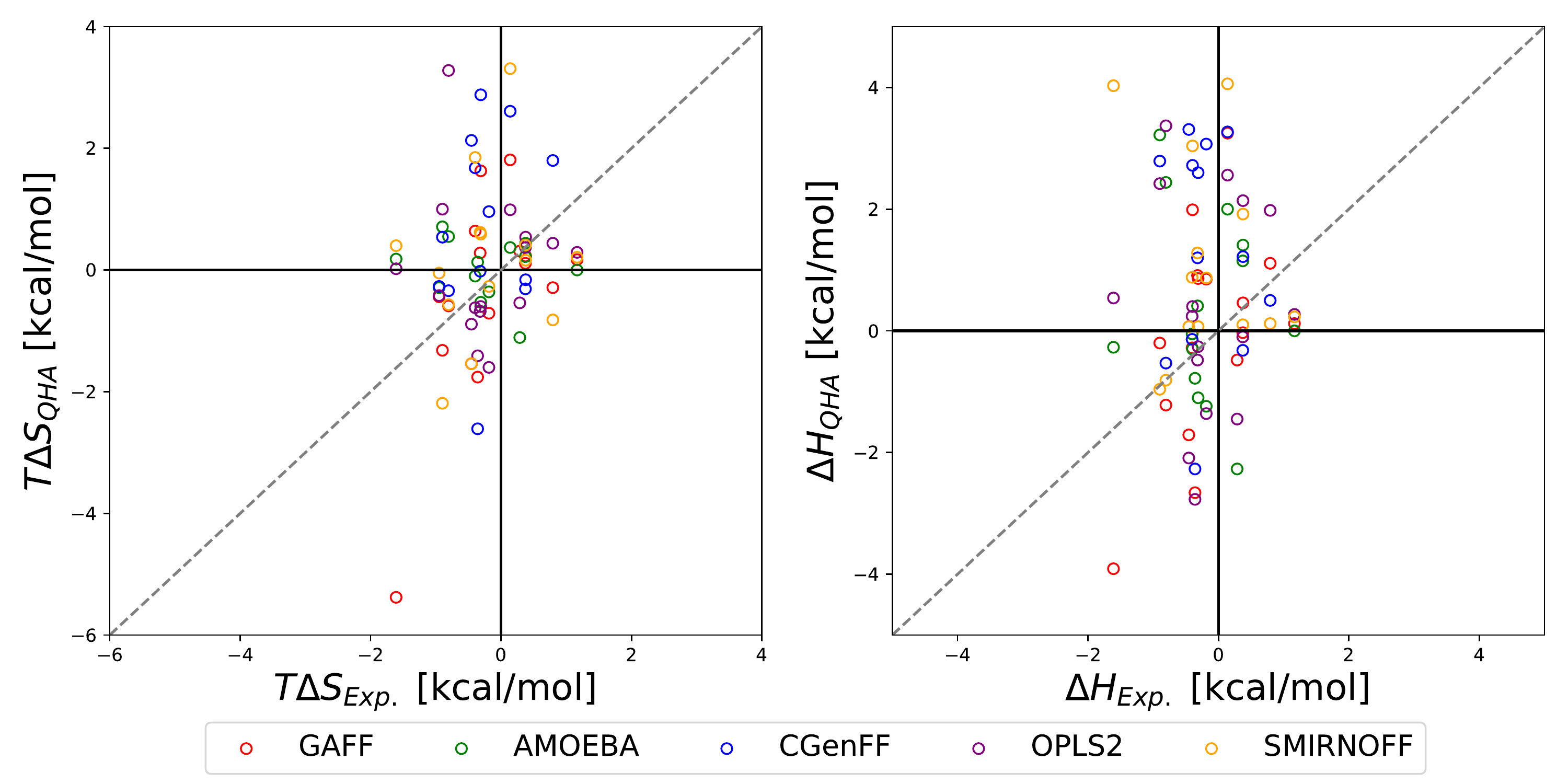} 
  \caption{Theoretical values of $T \Delta S$ and $\Delta H$ computed with QHA are plotted against the experimental values at the experimental transition temperature. The grey dashed line represents a 1-to-1 agreement between theoretical and experimental values. All force fields poorly predict the magnitude of the entropy and enthalpy contributions to the polymorph energy differences and their is poor agreement of all force fields with the experimental values.
  \label{figure:TdS_dH_scatter}}
\end{figure*}

    All five potentials poorly predict the magnitude of $\Delta H$ (RMSE$>$1.46 kcal/mol) and $T \Delta S$ (RMSE$>$0.93 kcal/mol) at the transition temperature with the entropy being predicted slightly better. Using the restructured lattice minima QHA/HA was performed using all five force fields. In some cases QHA would fail prior to the transition temperature so the free energy, entropy, and enthalpy differences were extrapolated with a linear fit of the last 50 K of the QHA data, providing the transition temperature thermodynamics reported in figure ~\ref{figure:TdS_dH_scatter}. The RMSE for each potential relative to experiment are reported in table ~\ref{table:RMSE_summary}, reflecting the error in the scatter plot where there is greater spread in the theoretical values than experiment. At the transition temperature $T \Delta S$ should equal $\Delta H$ and therefore the smaller RMSE for $T \Delta S$ implies that these force fields are more accurate at predicting the magnitude of the  entropy than enthalpy. The one exception to this is GAFF where the RMSE of $T \Delta S$ and $\Delta H$ are within bootstrapped error of one another. Plots for $\Delta G(T)$ for all sets of polymorphs are reported in the Supporting Information (section ~\ref{section:dGvT}).

    AMOEBA predicts the correct sign of $\Delta H$ for 71$\pm$12\% of the polymorphs and performs better than the fixed charge potentials, as expected. Getting the correct value of $\Delta H$ is important, but determining the correct sign will allow us to understand the correct low temperature stability of the polymorphs. Table ~\ref{table:RMSE_summary} reports the percentage for which the force field gets the correct sign of the polymorph entropy and enthalpy differences relative to experiment. If the sign of the enthalpy is left to random chance then there is a 50 \% chance of getting the correct sign. All 4 fixed charge potentials are within  a standard deviation of 50 \% and are not statistically different from random chance when determining the sign of $\Delta H$. This contrasts with AMOEBA, which gets the correct sign for 71$\pm$12\% of the polymorph pairs, performing better than random chance. This supports previous findings that polarizable potentials are better at determining potential energy differences than fixed charge potentials though the improvement is only moderate over the fixed charge potentials.

    Entropic differences are less sensitive to the force field than enthalpy, with all force fields performing statistically better than random chance except for CGenFF If the sign of the entropic difference between polymorphs is correctly computed, than we can determine the high temperature stability of the polymorphs. In table ~\ref{table:RMSE_summary} GAFF and OPLS2 correctly determine the sign of $T \Delta S$ for 75$\pm$11\% of the polymorph pairs statistically better than random chance. Both AMOEBA and SMIRNOFF perform slightly better than random chance, but more samples would improve the confidence in these force fields ability to determine the sign of the polymorph entropy differences.

\end{multicols}
\begin{center}
\begin{table}
\begin{tabular}{l||c|c||c|c} 
   \hline \hline
   & \multicolumn{2}{c||}{$T \Delta S$} & \multicolumn{2}{c}{$\Delta H$} \\ \hline
   & RMSE [kcal/mol] & Correct Sign & RMSE [kcal/mol] & Correct Sign \\ \hline \hline
  GAFF & 1.33$\pm$0.31 & 75$\pm$11\% & 1.46$\pm$0.24 & 62$\pm$12\% \\ \hline
  AMOEBA & 0.93$\pm$0.16 & 64$\pm$13\% & 1.79$\pm$0.36 & 71$\pm$12\% \\ \hline
  CGenFF & 1.72$\pm$0.26 & 46$\pm$14\% & 2.38$\pm$0.32 & 46$\pm$14\% \\ \hline
  OPLS2 & 1.35$\pm$0.38 & 75$\pm$11\% & 1.91$\pm$0.29 & 56$\pm$12\% \\ \hline
  SMIRNOFF & 1.42$\pm$0.25 & 64$\pm$13\% & 2.21$\pm$0.56 & 50$\pm$13\% \\ \hline \hline
\end{tabular}
\caption{The RMSE of the scatter plots in figure ~\ref{figure:TdS_dH_scatter} are reported for each force field relative to the experimental value at the transition temperature. Additionally, the percentage of polymorphs where the sign of enthalpy and entropy differences are correctly computed are reported. At the transition temperature the $T \Delta S = \Delta H$ and therefore the smaller RMSE in the entropy implies that all of the potentials, except GAFF, determine the magnitude of entropy differences with greater better than enthalpy. That being said, the RMSE of both entropy and enthalpy are approximately the same size or larger than the experimental values. Only AMOEBA performs better than random chance when determining the sign of $\Delta H$, which is contrasted with $T \Delta S$ where all potentials, except CGenFF, perform better than random chance. The error in the RMSE is the bootstrapped error.
\label{table:RMSE_summary}}
\end{table} 
\end{center}
\begin{multicols}{2}

    The signs of $\Delta H$ and $T \Delta S$ are correlated for all force fields except for CGenFF, allowing us to accurately model the enantiotropic behavior of 43--64$\pm$13\% of the polymorphs. Determining the correct sign in both entropy and enthalpy differences in polymorphs will allow us to determine both the high and low temperature stabilities of the polymorphs and therefore the change in ranking. Table ~\ref{table:sign_summary} breaks down the frequency for $\Delta H$ and $T \Delta S$ having the correct sign in relationship to one another. Based on random chance, the probability of any of the four categories in table ~\ref{table:sign_summary} should be 25 \%, any variation form 25 \% would indicate a correlation between the sign of the entropy and enthalpy. For all 4 force fields,  CGenFF excluded, the percentage of systems where the sign of $\Delta H$ and $T \Delta S$ are both correct with 43$\pm$13 -- 64$\pm$13 \% showing that the two properties are correlated.
 For GAFF, OPLS2, and AMOEBA we are able to correctly model the change in polymorph stability due to temperature for 57--64$\pm$13 \% of the polymorph pairs.

     
\end{multicols}
\begin{center}
\begin{table}
\begin{tabular}{l|c|c|c|c}
   \hline \hline
   & \multicolumn{4}{c}{Correct sign} \\ \hline 
   & $\Delta H$ \& $T \Delta S$ & $\Delta H$ & $T\Delta S$ & Neither \\ \hline \hline
  GAFF & 64$\pm$13\% & 7$\pm$7\% & 21$\pm$11\% & 21$\pm$11\% \\ \hline
  AMOEBA & 57$\pm$13\% & 14$\pm$9\% & 7$\pm$7\% & 21$\pm$11\% \\ \hline
  CGenFF & 36$\pm$13\% & 7$\pm$7\% & 7$\pm$7\% & 43$\pm$13\%  \\ \hline
  OPLS2 & 64$\pm$13\% & 0$\pm$0\% & 21$\pm$11\% & 29$\pm$12\% \\ \hline
  SMIRNOFF & 43$\pm$13\% & 7$\pm$7\% & 21$\pm$11\% & 29$\pm$12\% \\ \hline \hline
\end{tabular} 
\caption{For each force the percentage of systems where the sign of $\Delta H$ and $T \Delta S$ are both correct, both wrong, or independently correct are reported. The random chance of each category is 25 \%, so any percentage deviating by one standard deviation from that provides meaningful correlation. It is very likely for all force fields, except CGenFF, to get the correct sign of the entropy and enthalpy. 
\label{table:sign_summary}}
\end{table}
\end{center} 
\begin{multicols}{2}
%

\section{Conclusions}
    Fixed charge potentials perform on par with random probability (50 \%) for enthalpy differences, while the polarizable potential AMOEBA determines the correct sign for 71$\pm$12 \% of the polymorph pairs. Alternatively, all potentials except CGenFF determine the correct sign of entropy for 64$\pm$13 -- 75$\pm$11 \%, indicating that the entropy is less sensitive to the force field used. We also note that all force fields introduce error in  $\Delta H$ and $T \Delta  S$ (1--2.5 kcal/mol) at the transition temperature that is greater than or equal to the experimental values. The signs of of entropy and enthalpy differences are correlated  with one another, with 43--64 \% of the systems having both the entropy and enthalpy difference with the correct sign for all force fields except for CGenFF. The results here are limited in scope to a limited number of enantiotropic polymorphic transformations and require more systems for a statistically meaningful result.

\newpage
\newpage

\end{multicols}
\end{singlespace}

\begin{singlespace}
\begin{multicols}{2}
\bibliography{citations}
\end{multicols}
\end{singlespace}

\newpage

\maketitle

\section{Systems Studied} 
    In Table \ref{table:systems} we provide the systems studied, their polymorphs names/numberings, and corresponding Cambridge Structure Database reference code (CSD Refcode), and the dimensions of the supercells relative to the symmetric cell pulled from the database.
\singlespacing
      \begin{longtable}{l|l|l|l}
      \hline \hline
      \textbf{Piracetam} & Form I & Form III &\\
      CSD Refcode & BISMEV03 & BISMEV02 & \\
      Space Group & P2$_{1}$/n & P2$_{1}$/n &  \\
      Supercell Dimensions & $ 4\times 2\times 3$ & $ 2\times 3\times 4$ & \\
      \hline \hline
      

      \textbf{Carbamazepine} & Form I & Form III & Form IV \\
      CSD Refcode & CBMZPN11 & CBMZPN02 & \\
      Space Group & P$\bar{1}$ & P2$_{1}$/n &  \\
      Supercell Dimensions & $ 4\times 2\times 1$ & $ 4\times 2\times 2$ & $ \times \times $ \\
      \hline \hline

      \textbf{Adenine} & Form I & Form II & \\
      CSD Refcode & KOBFUD  & KOBFUD01 & \\
      Space Group & P2$_{1}$/c & Fdd2  &\\
      Supercell Dimensions & $ 3\times 1\times 4$ & $ 3\times 1\times 2$ & \\
      \hline \hline

      \textbf{Pyrazinamide} & $\alpha$ & $\beta$ & Form $\gamma$ \\
      CSD Refcode & PYRZIN14 & PYRZIN01 & PYRZIN20 \\
      Space Group & P2$_{1}$/a & P2$_{1}$/c & Pc \\
      Supercell Dimensions & $ 1\times 4\times 6$ & $ 2\times 6\times 2$ & $ 4\times 6\times 2$ \\
      \hline \hline

      \textbf{Resorcinol} & Form $\alpha$ & Form $\beta$ & \\
      CSD Refcode & RESORA03 & RESORA08 & \\
      Space Group & Pna2$_{1}$ & Pna2$_{1}$ & \\
      Supercell Dimensions & $ 2\times 2\times 4$ & $ 2\times 2\times 4$ & \\
      \hline \hline

      \textbf{Succinonitrile} & Form I & Form III & \\
      CSD Refcode & QOPBED  & N/A & \\
      Space Group & P2$_{1}$/a & N/A & \\
      Supercell Dimensions & $ 2\times 3\times 4$ & N/A & \\
      \hline \hline

      \textbf{Paracetamol} & Form I & Form II & \\
      CSD Refcode & HXACAN01 & HXACAN & \\
      Space Group & P2$_{1}$/a & Pcab & \\
      Supercell Dimensions & $ 2\times 2\times 3$ & $ 2\times 1\times 3$ & \\
      \hline \hline

      \textbf{Aripiprazole} & Form I & Form X & \\
      CSD Refcode & MELFIT01 & MELFIT05 & \\
      Space Group & P2$_{1}$ & P2$_{1}$ & \\
      Supercell Dimensions & $ 3\times 4\times 2$ & $ 3\times 4\times 2$ & \\
      \hline \hline

      \textbf{Tolbutamide} & Form I & Form II & Fomr III\\
      CSD Refcode & ZZZPUS04 & ZZZPUS05 & ZZZPUS09 \\
      Space Group & Pna2$_{1}$ & Pc & P2$_{1}$/n \\
      Supercell Dimensions & $ 2\times 3\times 2$ & $ 3\times 2\times 1$ & $ 2\times 3\times 2$\\
      \hline \hline

      \textbf{Chlorpropamide} & Form I & Form V & \\
      CSD Refcode & BEDMIG & BEDMIG04 & \\
      Space Group & P2$_{1}$2$_{1}$2$_{1}$ & Pna2$_{1}$ & \\
      Supercell Dimensions & $ 1\times 4\times 2$ & $ 1\times 4\times 2$ & \\
      \hline \hline
      \caption{Table of systems studied in the main paper. We have included the polymorph names, CSD reference codes, and the cell dimensions relative to the symmetric unit cell pulled down from the database.
      \label{table:systems}}
      \end{longtable}

\section{Free energy as a function of temperature} \label{section:dGvT}
    \begin{figure}
      \begin{center}
      \includegraphics[width=12cm]{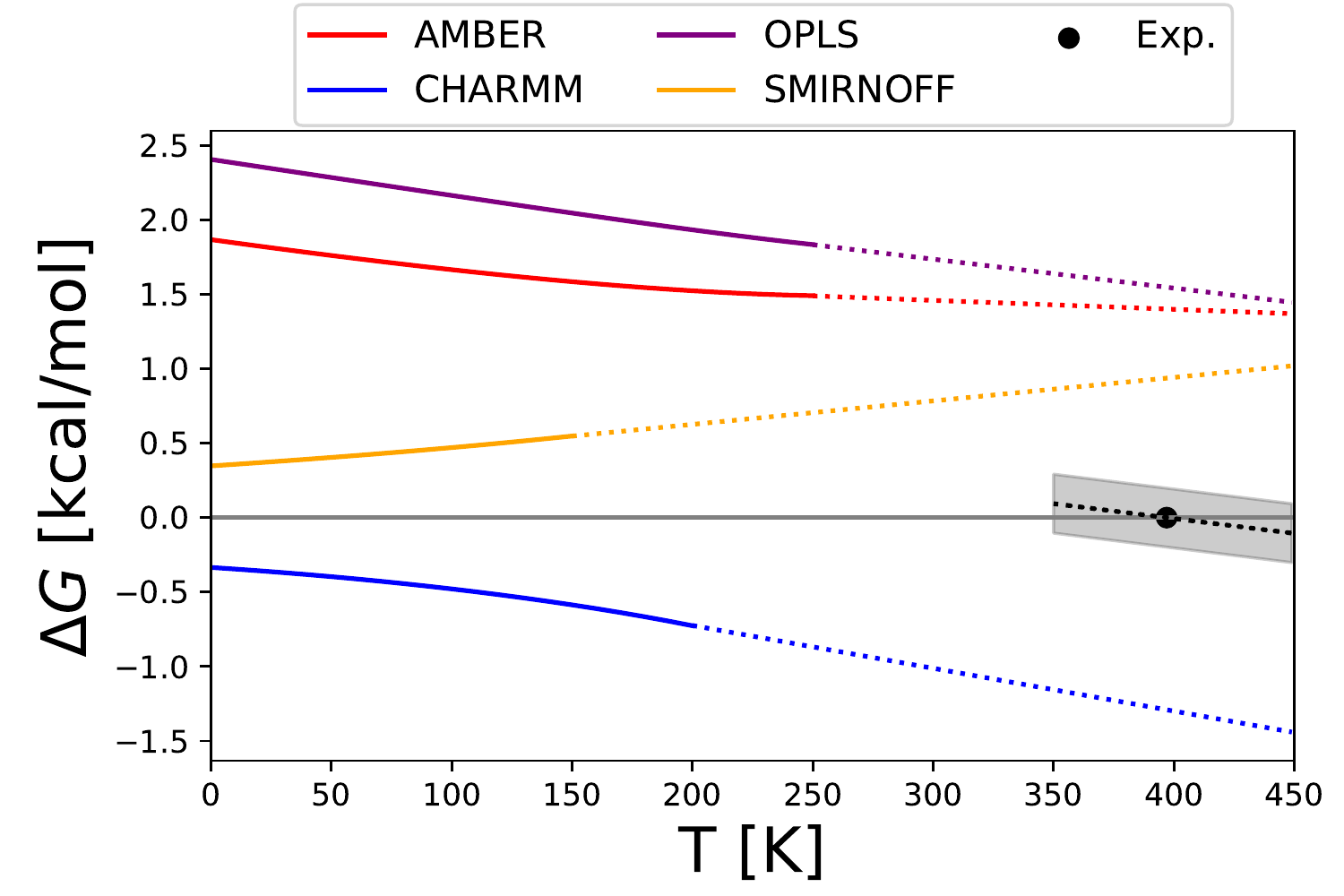}
      \end{center}
      \caption{Polymorph free energy difference of chlorproamide form V relative to I. Solid lines are directly computed from QHA  and the dashed lines are a linear fit of the last 50 K of $\Delta G(T)$. The experimental transition temperature is shown with a linear fit to $\Delta G = \Delta H - T\Delta S$ for the 100 K around the transition temperature.
      \label{figure:}}
    \end{figure}

    \begin{figure}
      \begin{center}
      \includegraphics[width=12cm]{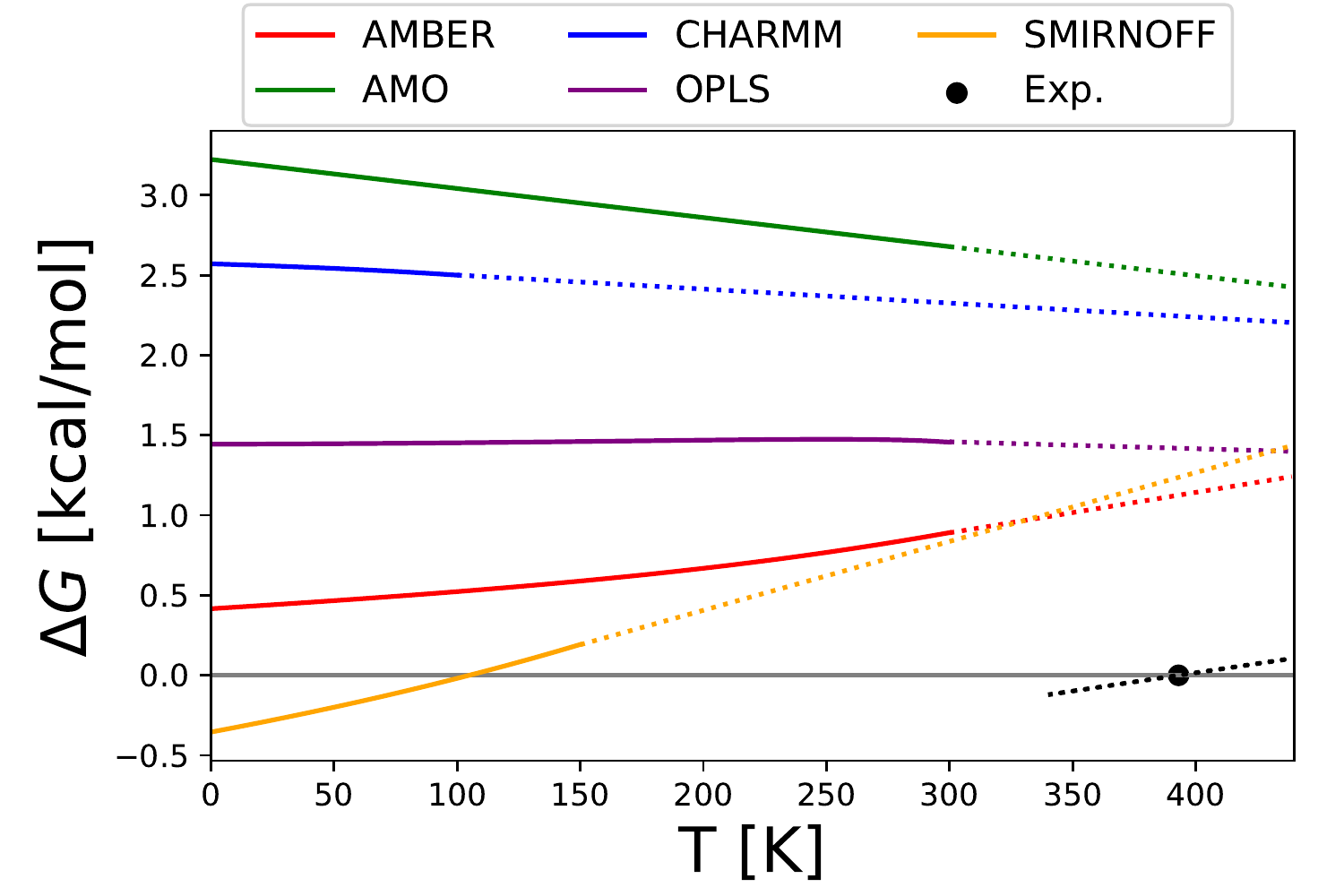}
      \end{center}
      \caption{Polymorph free energy difference of piracetam form III relative to I. Solid lines are directly computed from QHA  and the dashed lines are a linear fit of the last 50 K of $\Delta G(T)$. The experimental transition temperature is shown with a linear fit to $\Delta G = \Delta H - T\Delta S$ for the 100 K around the transition temperature.
      \label{figure:}}
    \end{figure}

    \begin{figure}
      \begin{center}
      \includegraphics[width=12cm]{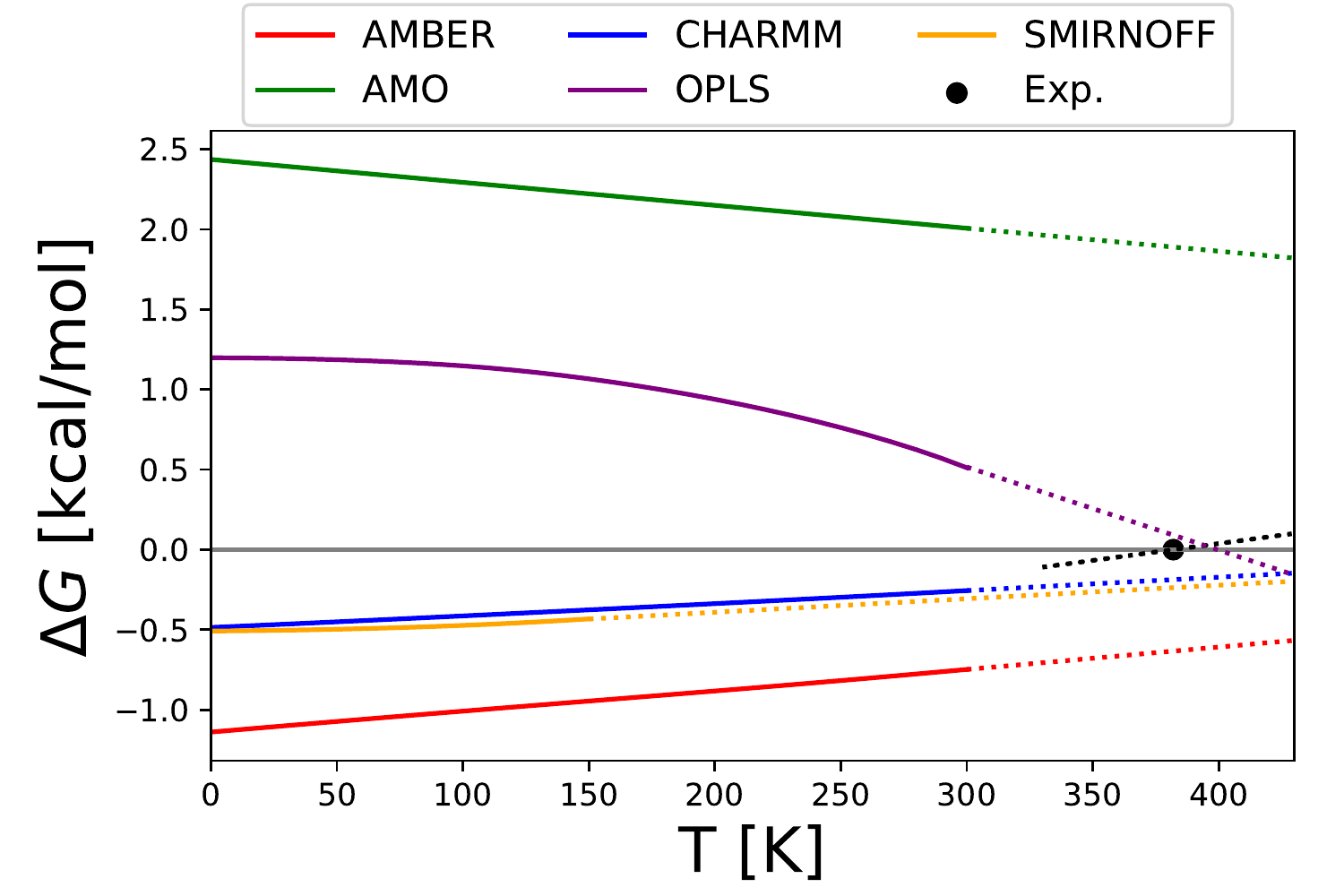}
      \end{center}
      \caption{Polymorph free energy difference of piracetam form II relative to I. Solid lines are directly computed from QHA  and the dashed lines are a linear fit of the last 50 K of $\Delta G(T)$. The experimental transition temperature is shown with a linear fit to $\Delta G = \Delta H - T\Delta S$ for the 100 K around the transition temperature.
      \label{figure:}}
    \end{figure}

    \begin{figure}
      \begin{center}
      \includegraphics[width=12cm]{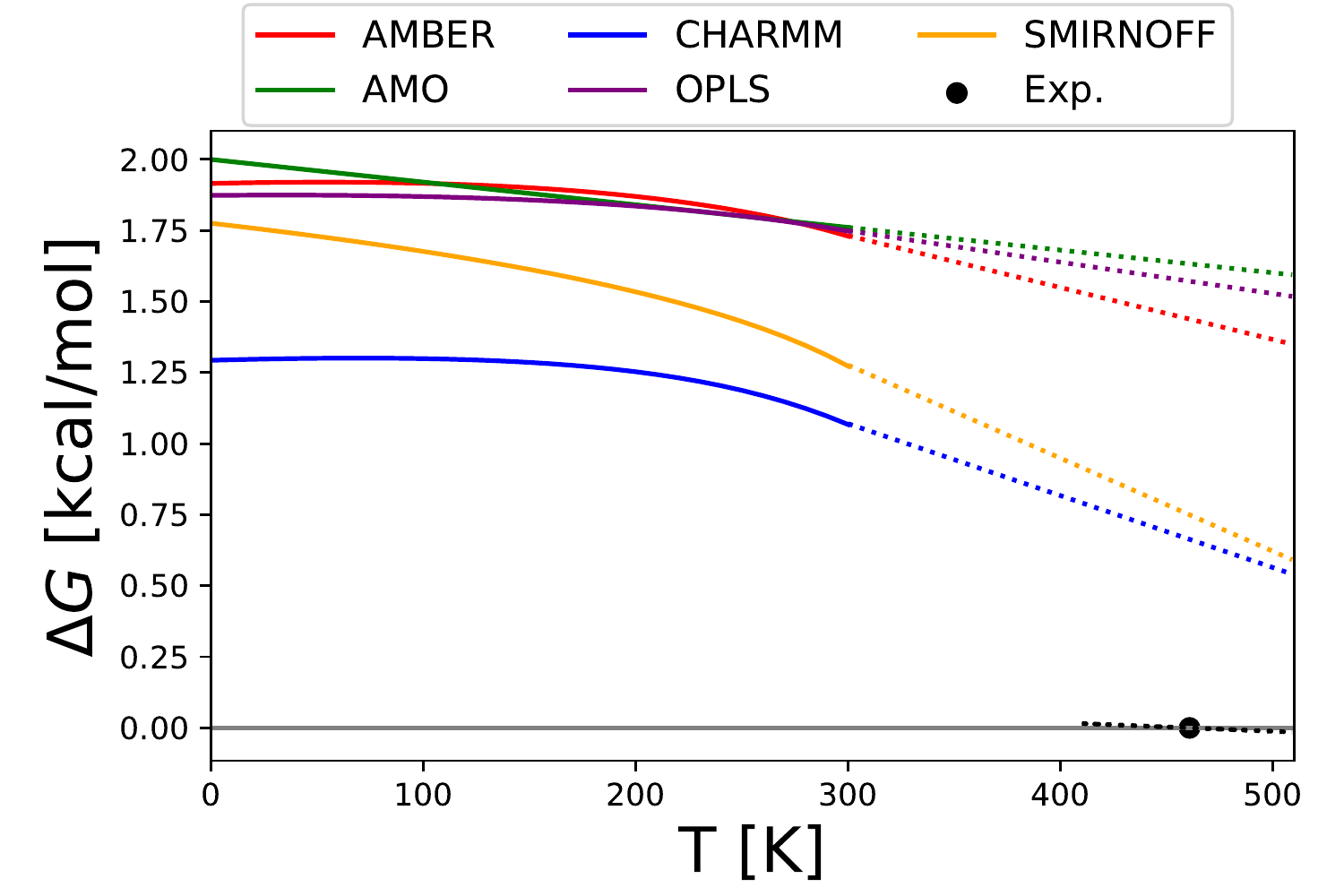}
      \end{center}
      \caption{Polymorph free energy difference of carbamzepine form IV relative to III. Solid lines are directly computed from QHA  and the dashed lines are a linear fit of the last 50 K of $\Delta G(T)$. The experimental transition temperature is shown with a linear fit to $\Delta G = \Delta H - T\Delta S$ for the 100 K around the transition temperature.
      \label{figure:}}
    \end{figure}

    \begin{figure}
      \begin{center}
      \includegraphics[width=12cm]{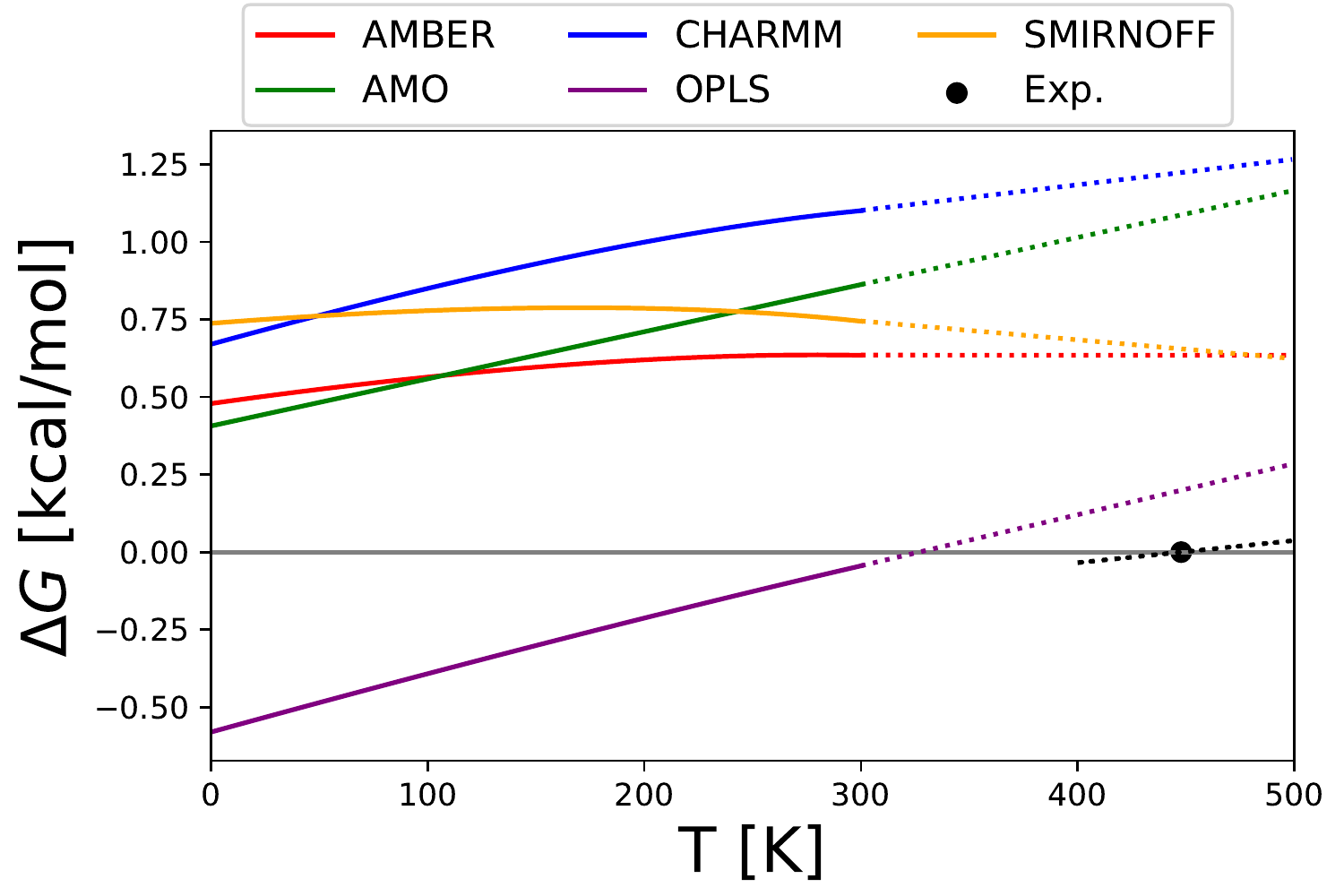}
      \end{center}
      \caption{Polymorph free energy difference of carbamzepine form I relative to III. Solid lines are directly computed from QHA  and the dashed lines are a linear fit of the last 50 K of $\Delta G(T)$. The experimental transition temperature is shown with a linear fit to $\Delta G = \Delta H - T\Delta S$ for the 100 K around the transition temperature.
      \label{figure:}}
    \end{figure}

    \begin{figure}
      \begin{center}
      \includegraphics[width=12cm]{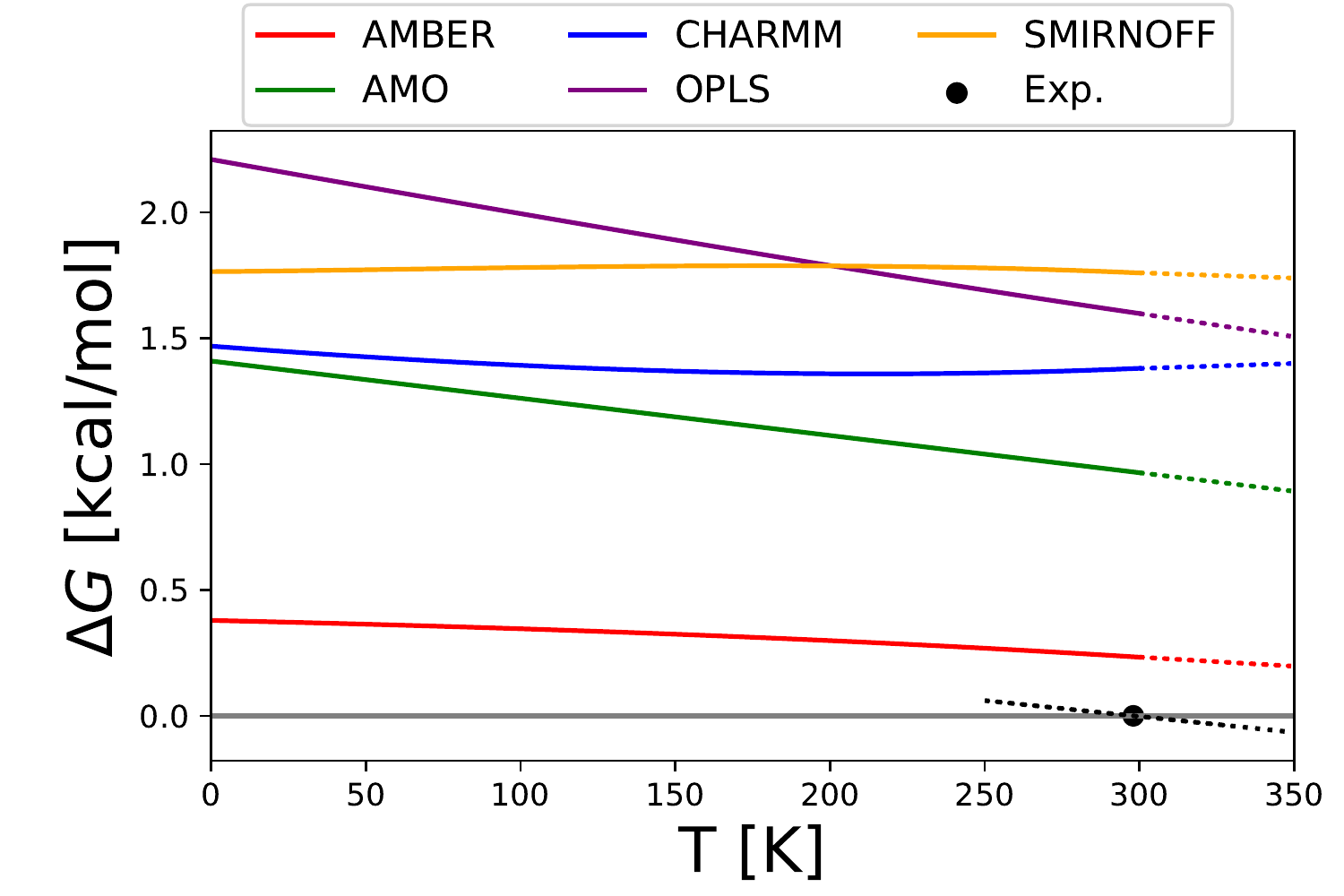}
      \end{center}
      \caption{Polymorph free energy difference of adenine form II relative to I. Solid lines are directly computed from QHA  and the dashed lines are a linear fit of the last 50 K of $\Delta G(T)$. The experimental transition temperature is shown with a linear fit to $\Delta G = \Delta H - T\Delta S$ for the 100 K around the transition temperature.
      \label{figure:}}
    \end{figure}

    \begin{figure}
      \begin{center}
      \includegraphics[width=12cm]{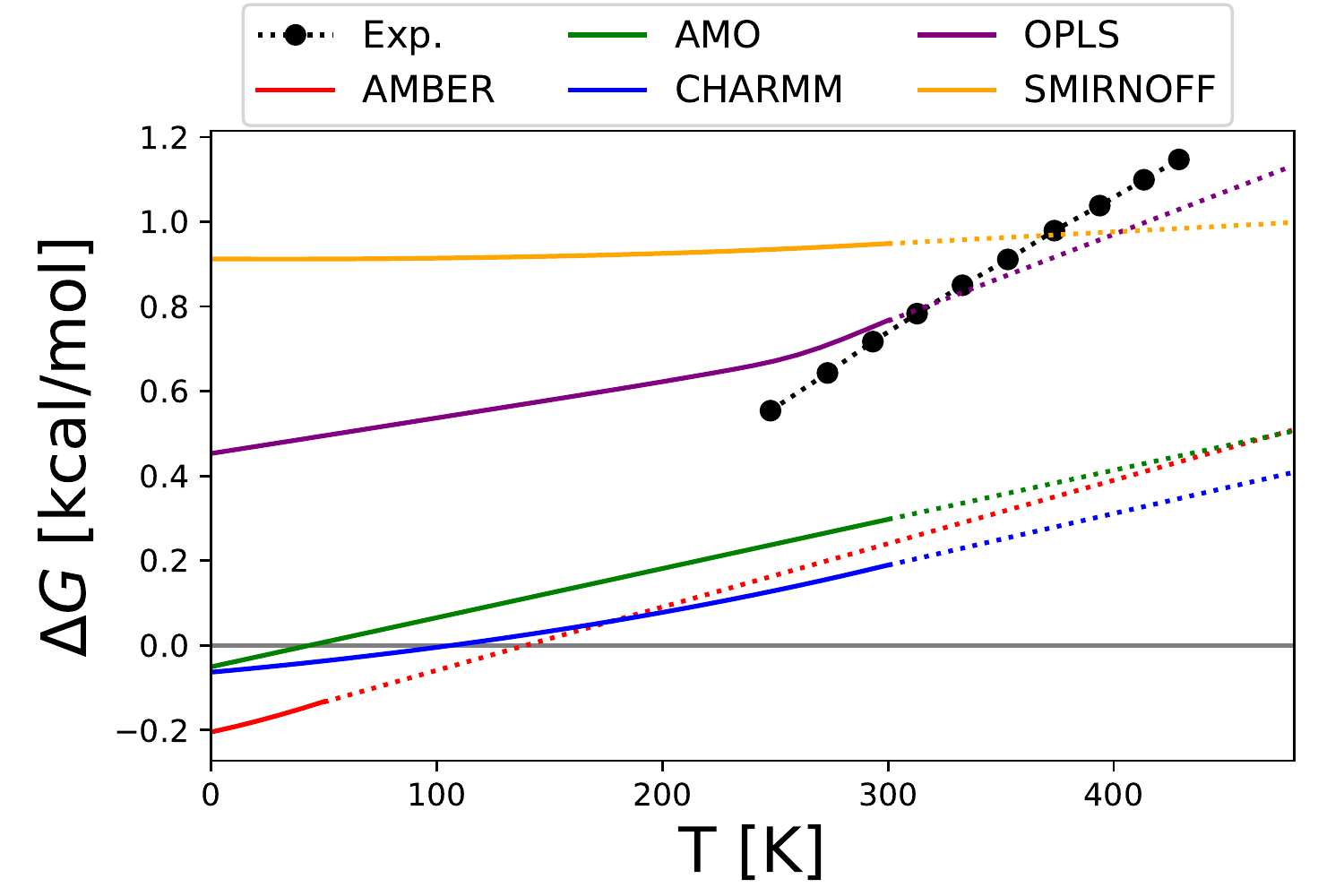}
      \end{center}
      \caption{Polymorph free energy difference of paracetamol form I relative to II. Solid lines are directly computed from QHA  and the dashed lines are a linear fit of the last 50 K of $\Delta G(T)$. The experimental transition temperature is shown with a linear fit to $\Delta G = \Delta H - T\Delta S$ for the 100 K around the transition temperature.
      \label{figure:}}
    \end{figure}

    \begin{figure}
      \begin{center}
      \includegraphics[width=12cm]{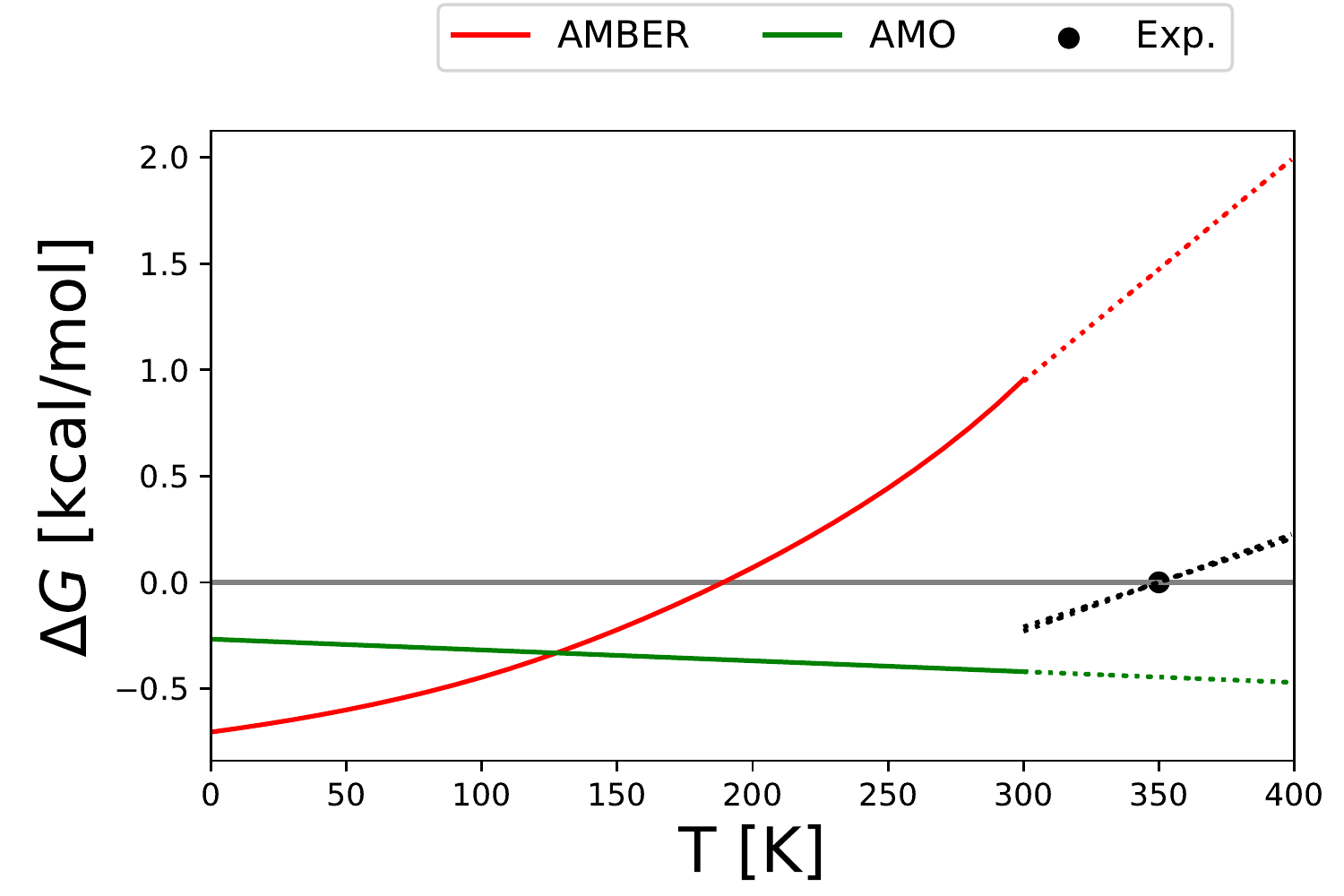}
      \end{center}
      \caption{Polymorph free energy difference of aripiprazole form X relative to I. Solid lines are directly computed from QHA  and the dashed lines are a linear fit of the last 50 K of $\Delta G(T)$. The experimental transition temperature is shown with a linear fit to $\Delta G = \Delta H - T\Delta S$ for the 100 K around the transition temperature.
      \label{figure:}}
    \end{figure}

    \begin{figure}
      \begin{center}
      \includegraphics[width=12cm]{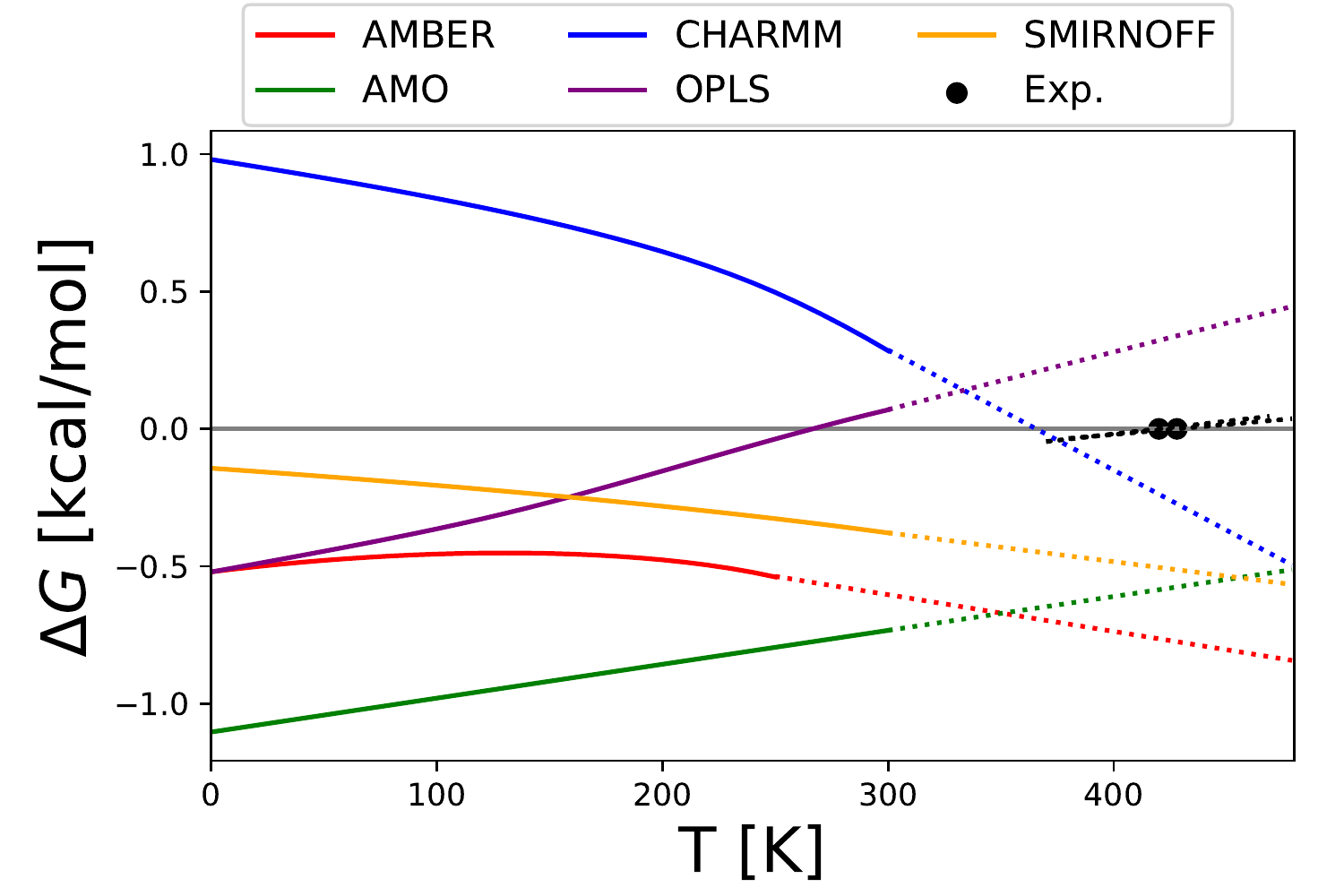}
      \end{center}
      \caption{Polymorph free energy difference of pyrzinamide form $\alpha$ relative to $\gamma$. Solid lines are directly computed from QHA  and the dashed lines are a linear fit of the last 50 K of $\Delta G(T)$. The experimental transition temperature is shown with a linear fit to $\Delta G = \Delta H - T\Delta S$ for the 100 K around the transition temperature.
      \label{figure:}}
    \end{figure}

    \begin{figure}
      \begin{center}
      \includegraphics[width=12cm]{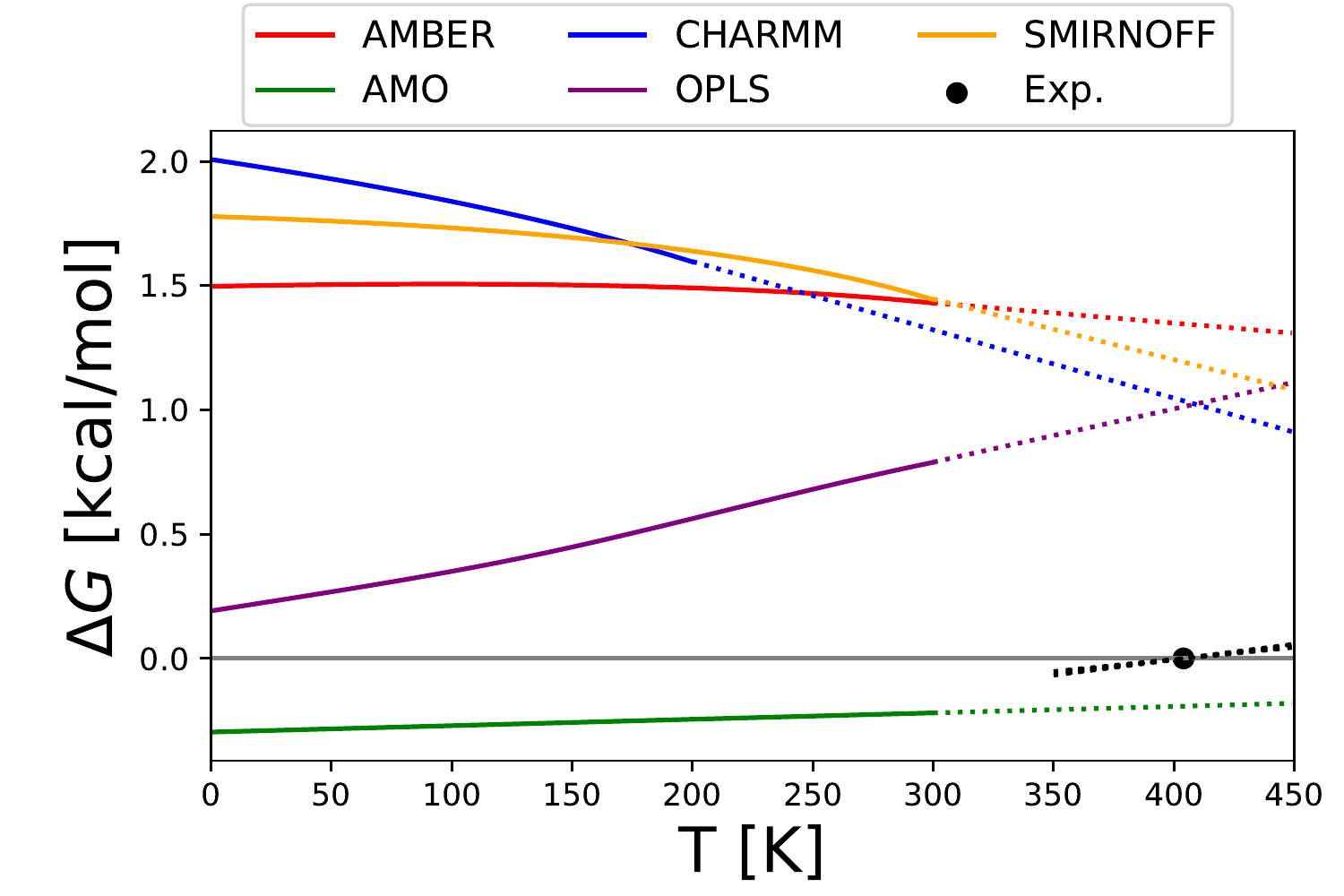}
      \end{center}
      \caption{Polymorph free energy difference of pyrzinamide form $\delta$ relative to $\gamma$. Solid lines are directly computed from QHA  and the dashed lines are a linear fit of the last 50 K of $\Delta G(T)$. The experimental transition temperature is shown with a linear fit to $\Delta G = \Delta H - T\Delta S$ for the 100 K around the transition temperature.
      \label{figure:}}
    \end{figure}

    \begin{figure}
      \begin{center}
      \includegraphics[width=12cm]{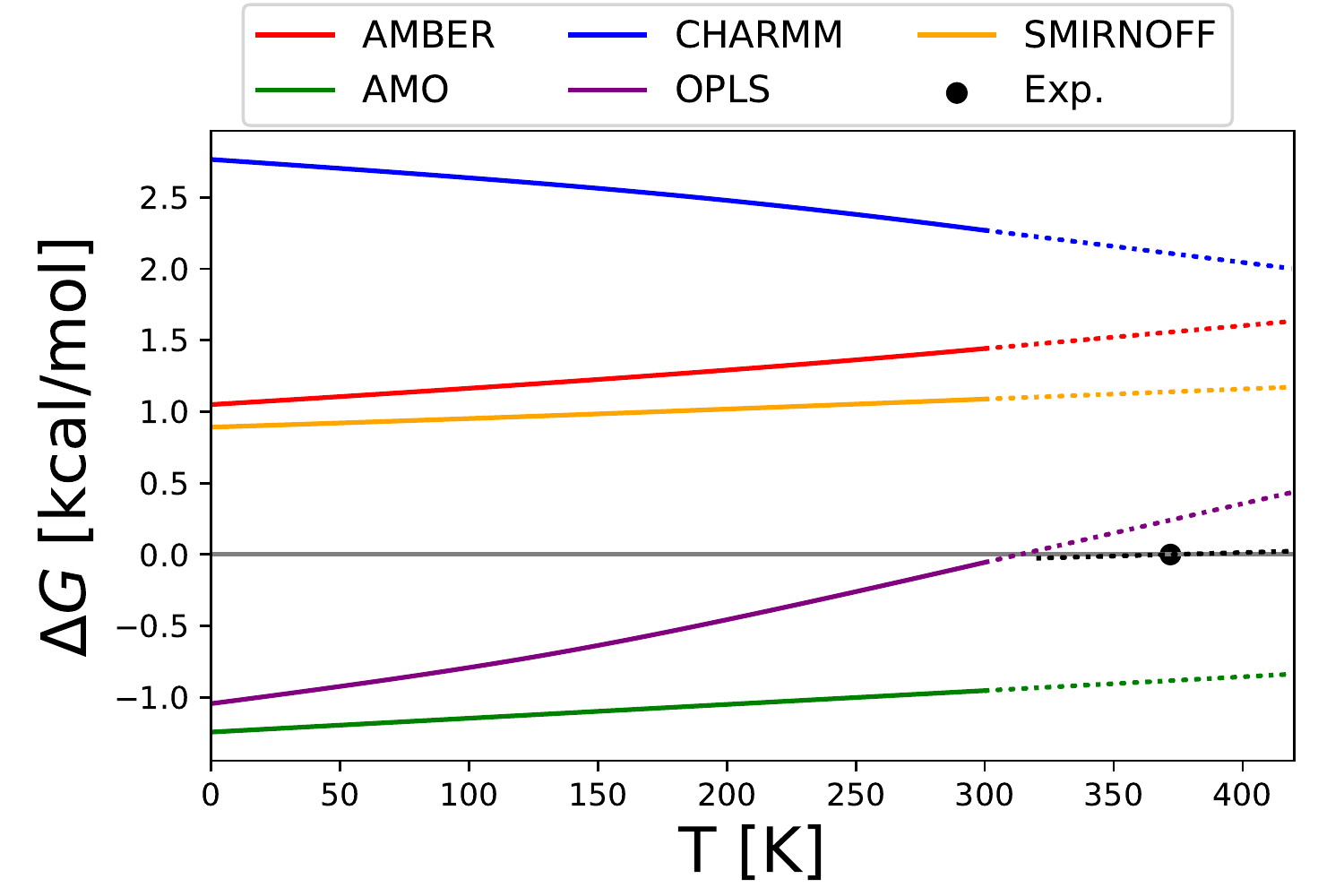}
      \end{center}
      \caption{Polymorph free energy difference of pyrzinamide form $\beta$ relative to $\gamma$. Solid lines are directly computed from QHA  and the dashed lines are a linear fit of the last 50 K of $\Delta G(T)$. The experimental transition temperature is shown with a linear fit to $\Delta G = \Delta H - T\Delta S$ for the 100 K around the transition temperature.
      \label{figure:}}
    \end{figure}

    \begin{figure}
      \begin{center}
      \includegraphics[width=12cm]{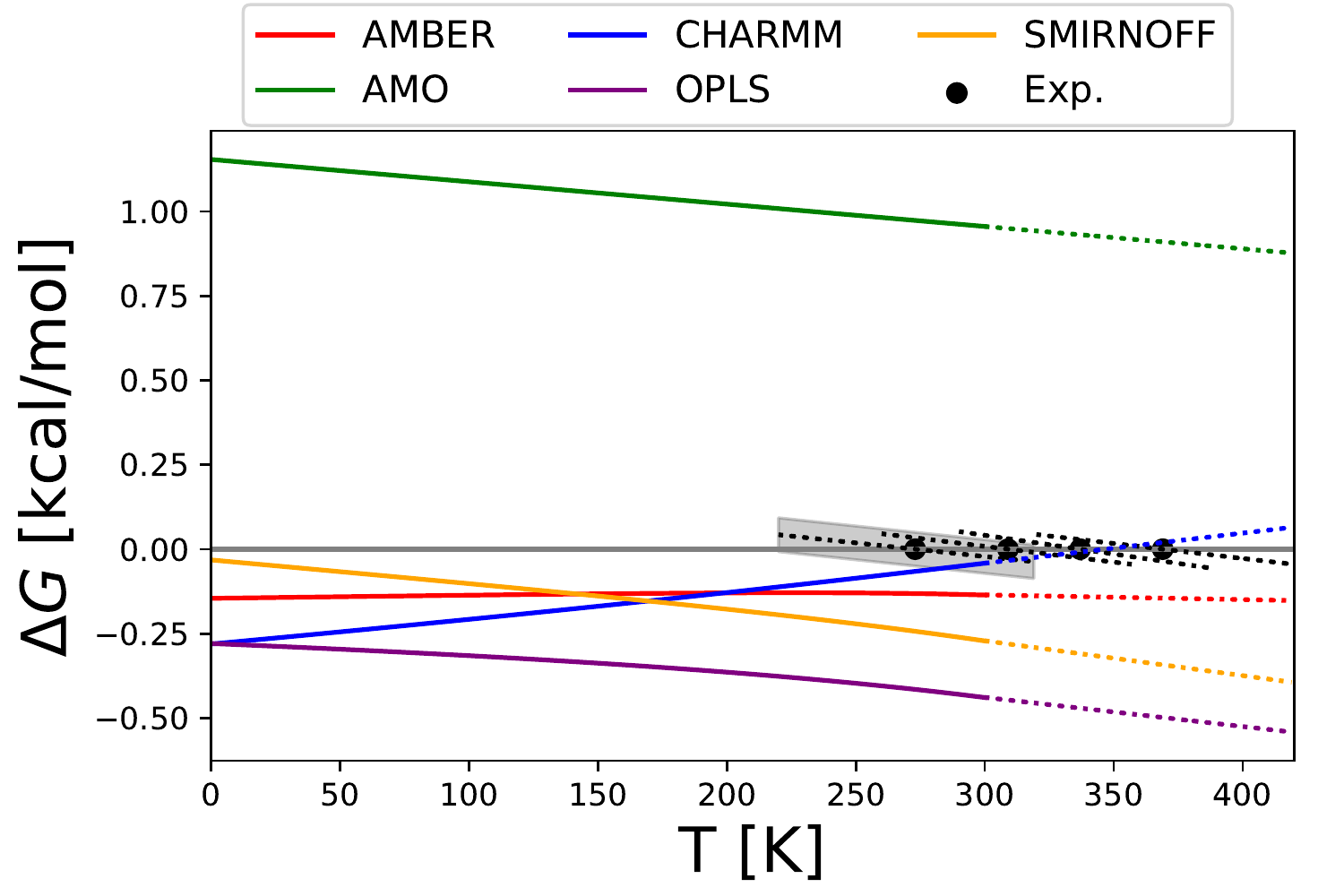}
      \end{center}
      \caption{Polymorph free energy difference of resorcinol form $\beta$ relative to $\alpha$. Solid lines are directly computed from QHA  and the dashed lines are a linear fit of the last 50 K of $\Delta G(T)$. The experimental transition temperature is shown with a linear fit to $\Delta G = \Delta H - T\Delta S$ for the 100 K around the transition temperature.
      \label{figure:}}
    \end{figure}
    
    \begin{figure}
      \begin{center}
      \includegraphics[width=12cm]{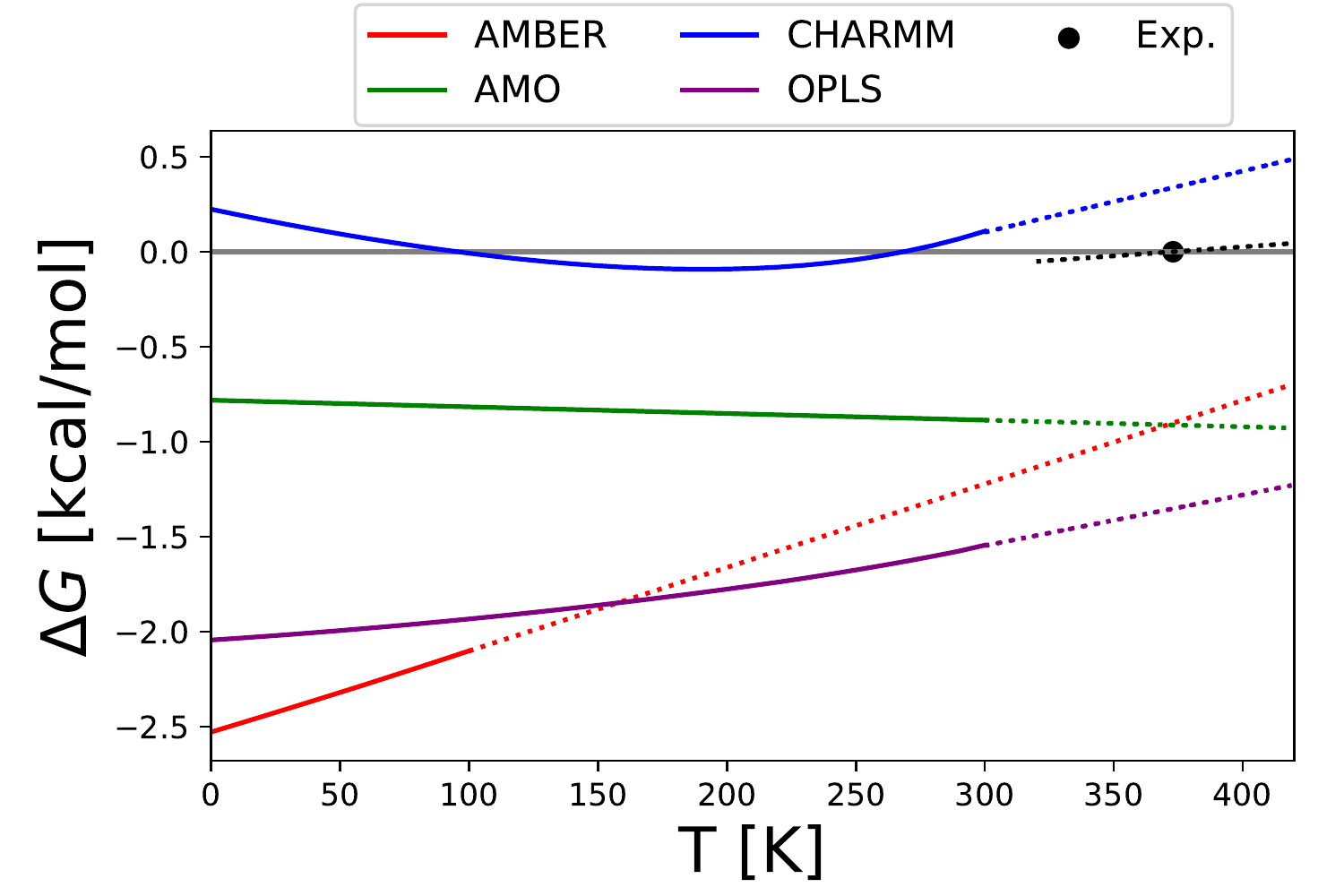}
      \end{center}
      \caption{Polymorph free energy difference of tolbutamide form II relative to I. Solid lines are directly computed from QHA  and the dashed lines are a linear fit of the last 50 K of $\Delta G(T)$. The experimental transition temperature is shown with a linear fit to $\Delta G = \Delta H - T\Delta S$ for the 100 K around the transition temperature.
      \label{figure:}}
    \end{figure}

    \begin{figure}
      \begin{center}
      \includegraphics[width=12cm]{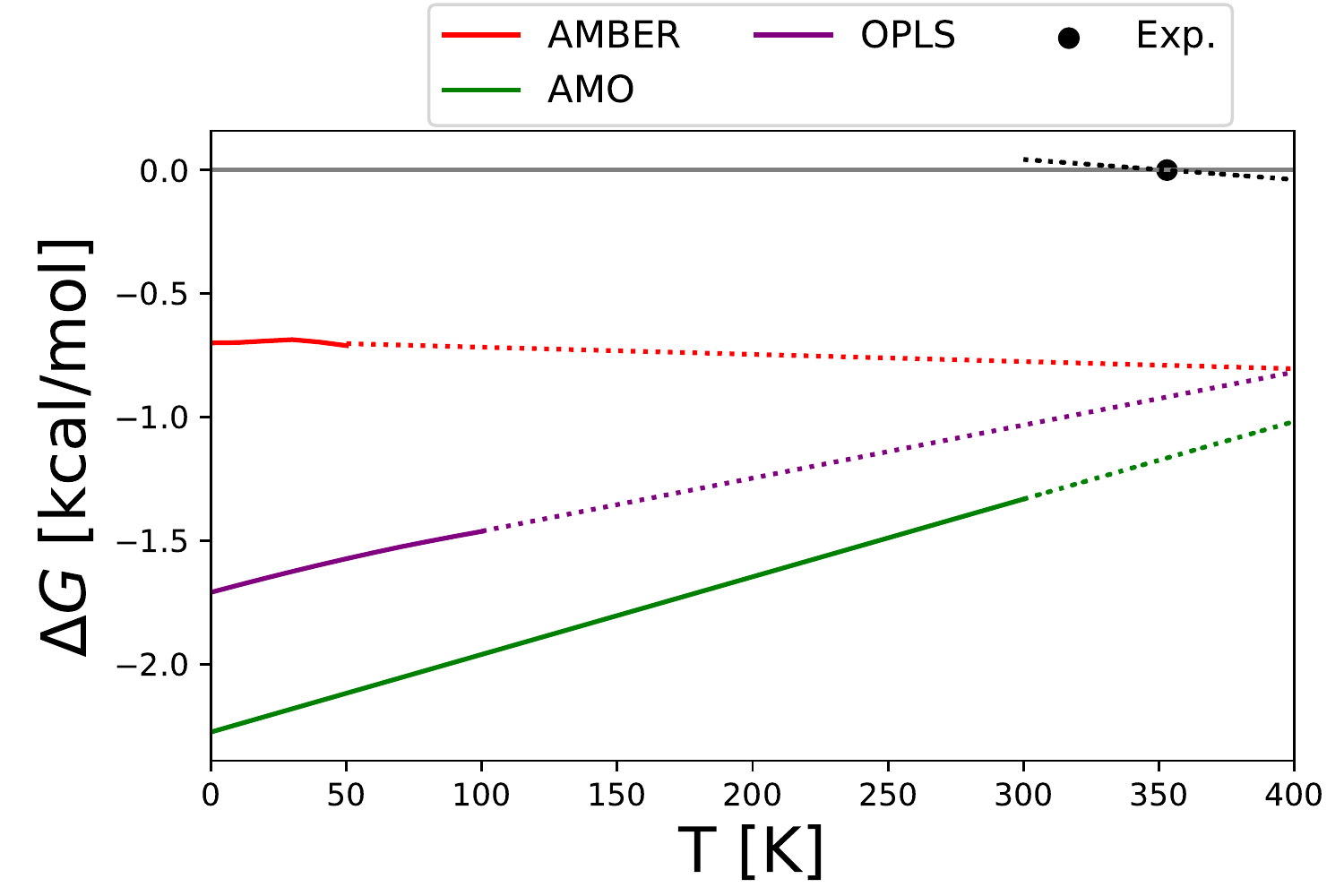}
      \end{center}
      \caption{Polymorph free energy difference of tolbutamide form IV relative to I. Solid lines are directly computed from QHA  and the dashed lines are a linear fit of the last 50 K of $\Delta G(T)$. The experimental transition temperature is shown with a linear fit to $\Delta G = \Delta H - T\Delta S$ for the 100 K around the transition temperature.
      \label{figure:}}
    \end{figure}

    \begin{figure}
      \begin{center}
      \includegraphics[width=12cm]{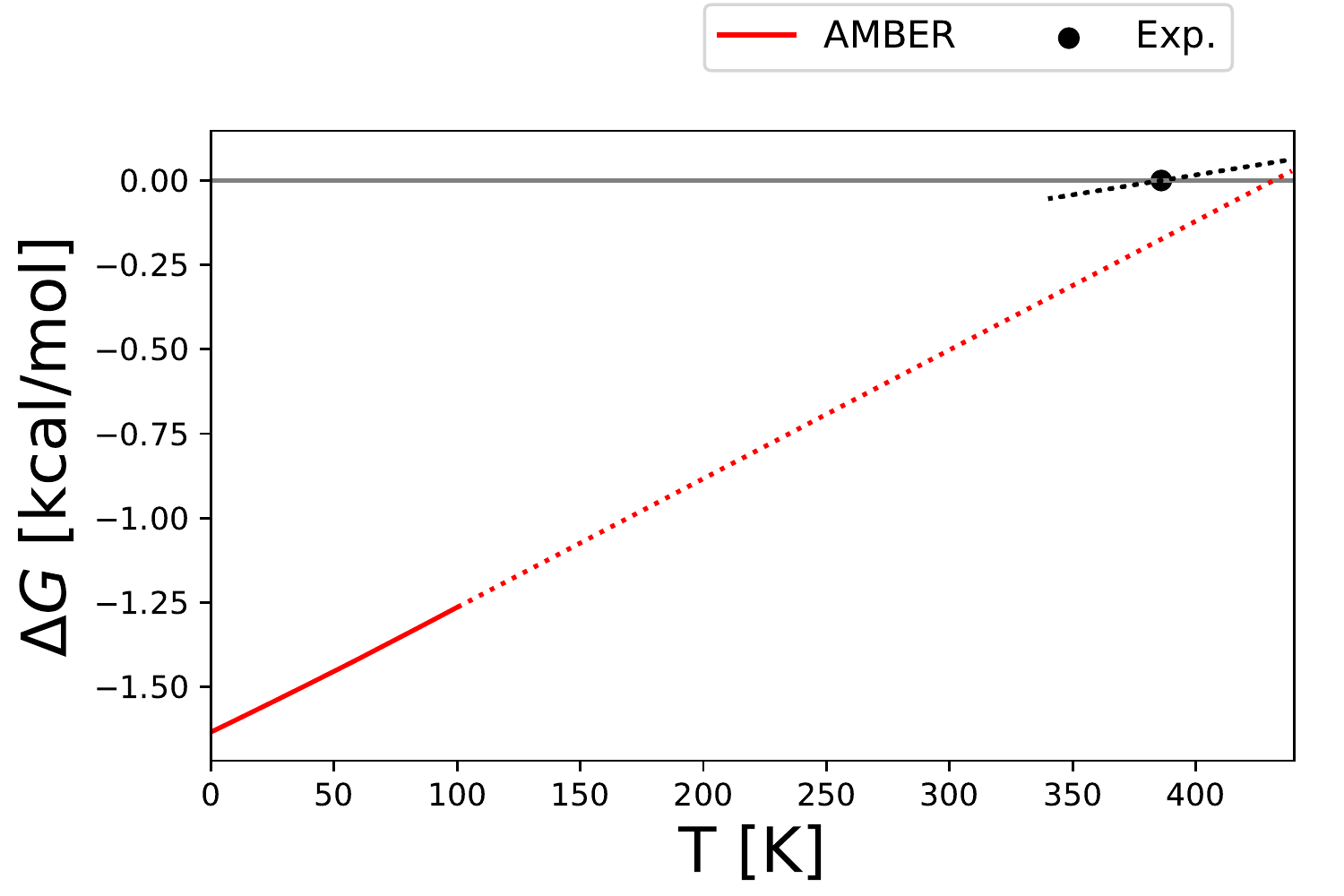}
      \end{center}
      \caption{Polymorph free energy difference of tolbutamide form III relative to I. Solid lines are directly computed from QHA  and the dashed lines are a linear fit of the last 50 K of $\Delta G(T)$. The experimental transition temperature is shown with a linear fit to $\Delta G = \Delta H - T\Delta S$ for the 100 K around the transition temperature.
      \label{figure:}}
    \end{figure}

    \begin{figure}
      \begin{center}
      \includegraphics[width=12cm]{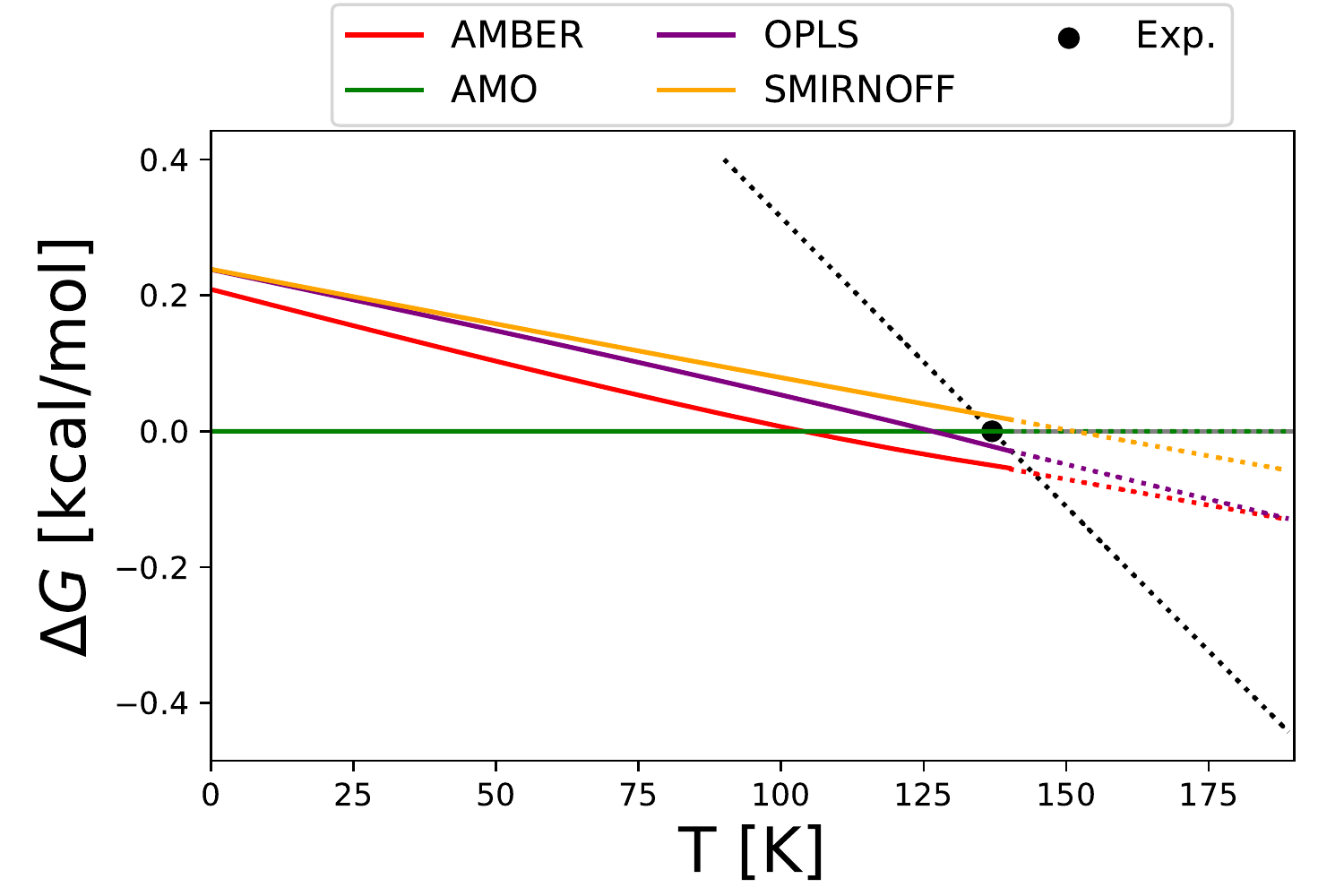}
      \end{center}
      \caption{Polymorph free energy difference of cyclopentane form I relative to III. Solid lines are directly computed from QHA  and the dashed lines are a linear fit of the last 50 K of $\Delta G(T)$. The experimental transition temperature is shown with a linear fit to $\Delta G = \Delta H - T\Delta S$ for the 100 K around the transition temperature.
      \label{figure:}}
    \end{figure}

\newpage

\section{Summary of energy differences}
\begin{figure*}
  \includegraphics[width=16cm]{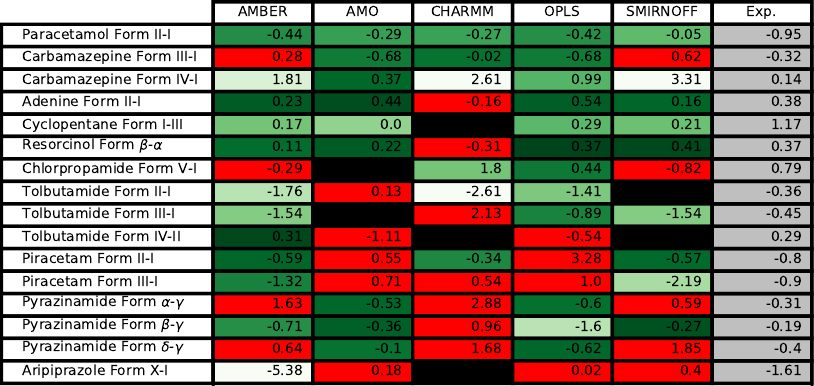} \\ 
  \includegraphics[width=8cm]{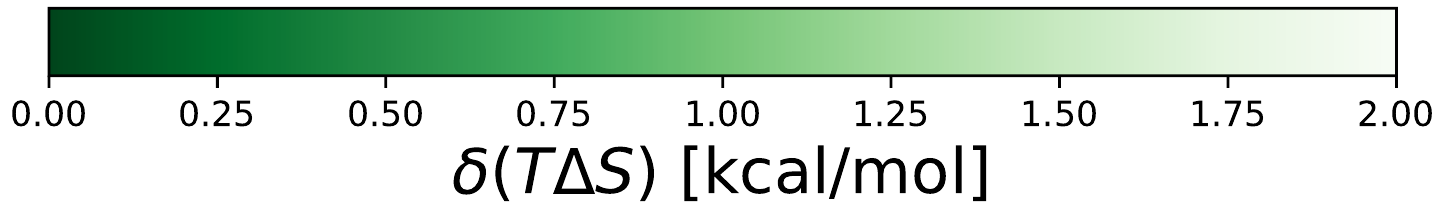}
  \caption{$T \Delta S$ for the 5 classical potentials and the experimental value at the transition temperature. Cells blacked out indicate that a lattice minimum could not be found to run QHA from. Red cells indicate that the value of $T \Delta S$ is the incorrect sign. The remaining cells are colored based on the error from the experimental value. AMBER and OPLS perform the best at getting the correct sign of the entropy with CHARMM performing the worst.
  \label{table:TdS_table}}
\end{figure*}
\begin{figure*}
  \includegraphics[width=16cm]{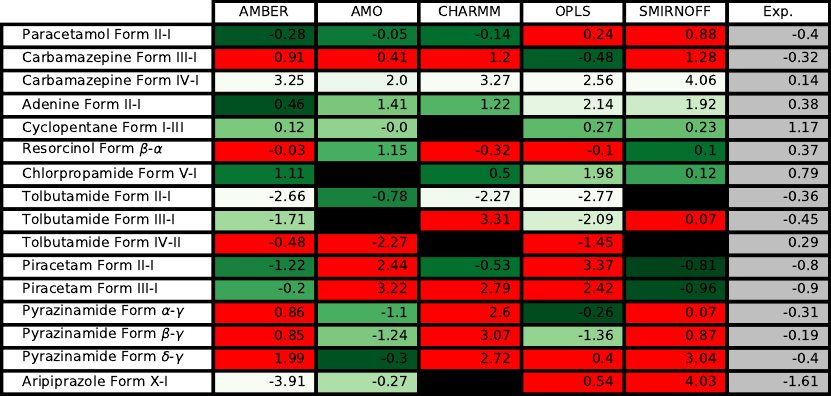} \\ 
  \includegraphics[width=8cm]{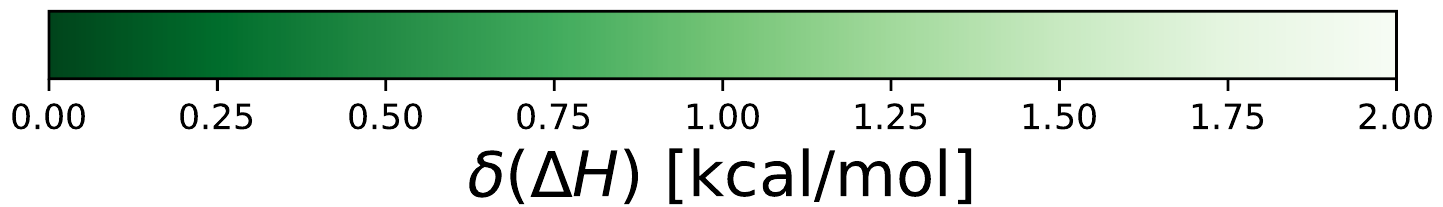}
  \caption{$\Delta H$ for the 5 classical potentials and the experimental value at the transition temperature. Cells blacked out indicate that a lattice minimum could not be found to run QHA from. Red cells indicate that the value of $\Delta H$ is the incorrect sign. The remaining cells are colored based on the error from the experimental value. AMBER and AMOEBA perform the best at getting the correct sign, but overall the results here show that the entropy is determined better with the classical potentials than the enthalpy.
  \label{table:dH_table}}
\end{figure*}
%

\newpage
\newpage

\end{document}